\documentclass{article}

\usepackage[preprint]{neurips_2026}

\usepackage[utf8]{inputenc} 
\usepackage[T1]{fontenc}    
\usepackage{hyperref}       
\usepackage{url}            
\usepackage{booktabs}       
\usepackage{amsfonts}       
\usepackage{nicefrac}       
\usepackage{microtype}      
\usepackage{xcolor}         
\usepackage{amsmath,amssymb}
\usepackage{graphicx}
\usepackage{cleveref}
\usepackage{algorithm}
\usepackage{algpseudocodex}
\usepackage{subcaption}
\usepackage{multirow}
\usepackage{enumitem}
\usepackage{xspace}
\usepackage[skins,breakable]{tcolorbox}
\usepackage{wrapfig}
\usepackage{graphicx} 

\newcommand{\ours}{\textsc{CoHarden}\xspace}
\newcommand{\ftp}{\texorpdfstring{F$\to$P}{F->P}\xspace}

\newcommand{\tdd}{\textsc{TDD}\xspace}
\newcommand{\fixonly}{\textsc{FIX-only}\xspace}
\newcommand{\brtonly}{\textsc{BRT-only}\xspace}
\newcommand{\cogen}{\textsc{Cogen}\xspace}
\newcommand{\mpe}{\textsc{MPE}\xspace}

\newcommand{\eat}[1]{}

\crefname{algocf}{Algorithm}{Algorithms}
\Crefname{algocf}{Algorithm}{Algorithms}

\newtcolorbox{rqbox}[1][]{%
  enhanced, breakable,
  colback=gray!5, colframe=black!55,
  boxrule=0.5pt, arc=2pt,
  left=6pt, right=6pt, top=4pt, bottom=4pt,
  fonttitle=\bfseries,
  #1}

\title{Beyond Fail-to-Pass: Iterative Hardening of Co-Generated Bug Reproduction Tests and Fixes}

\author{%
  \textbf{Yuhao Tan}$^{1,3}$ \quad \textbf{Zhibang Yang}$^{2,3}$ \quad \textbf{Fangkai Yang}$^{3\dagger}$ \quad \textbf{Yuan Yao}$^{1}$ \quad \textbf{Yu Kang}$^{3}$ \\[0.2em]
  \textbf{Lu Wang}$^{3}$ \quad \textbf{Pu Zhao}$^{3}$ \quad \textbf{Xin Zhang}$^{3}$ \quad \textbf{Xiaoxing Ma}$^{1\dagger}$ \quad \textbf{Qingwei Lin}$^{3}$ \\[0.2em]
  \textbf{Saravan Rajmohan}$^{3}$ \quad \textbf{Dongmei Zhang}$^{3}$ \\[0.5em]
  $^{1}$Nanjing University \quad $^{2}$Peking University \quad $^{3}$Microsoft \\[0.3em]
  \small $^{\dagger}$Corresponding authors: \texttt{fangkaiyang@microsoft.com}, \texttt{xxm@nju.edu.cn} \\[0.2em]
  \small\texttt{yhtan@smail.nju.edu.cn} \\[0.2em]
  \small \textbf{Project page}: \url{https://aka.ms/CoHarden}
}

\begin{document}

\maketitle


\begin{abstract}
Large language models (LLMs) have made automated program repair (APR) increasingly practical for real-world bugs, but repairing directly from bug reports remains underconstrained. 
Bug reproduction tests (BRTs) help close this gap by turning a bug report into an executable, bug-specific signal that can guide repair and validate candidate patches. Existing work has therefore studied BRT generation as a core subproblem in APR and mainly evaluates a generated BRT using the fail-to-pass (\ftp) criterion, which requires the test to fail on the buggy code but pass on the golden fix.
We show that \ftp alone is insufficient when the goal of a BRT is to improve downstream repair. In particular, some \ftp BRTs are \emph{lax}, reproducing the observed symptom yet still admitting plausible-but-incorrect patches. We formalize this missing quality dimension by separating \ftp BRTs into \emph{rigorous} and \emph{lax} ones, and show empirically that only the former consistently improve repair success. We further find that co-generation introduces test--fix error coupling, where the in-trajectory fail-to-pass (\ftp) check can pass even when both the generated patch and generated test are wrong.
Based on these findings, we propose \ours{}, a co-generation framework that uses the Lax signal as an in-loop convergence criterion. \ours{} first generates a test before any fix, then iteratively hardens the test and fix against surviving mutation patches until the generated test no longer admits Lax regressions. Experiments show that \ours{} reaches $69.4\%$ Resolved and $78.9\%$ \ftp{} on SWE-bench Verified, outperforming the strongest fix-only and cogeneration baselines by $+9.6$ and $+7.9$ percentage points in Resolved, respectively, with consistent gains across LLM backbones and benchmarks.

\end{abstract}

\section{Introduction}
\label{sec:intro}



Automated Program Repair (APR) has advanced rapidly with Large Language Models (LLMs), and recent agentic systems can generate increasingly plausible fixes for real-world software bugs~\citep{DBLP:conf/icse/XiaWZ23,DBLP:conf/nips/YangJWLYNP24,DBLP:conf/issta/0002RFR24,DBLP:journals/pacmse/XiaDDZ25}. However, a plausible fix is not the same as a correct fix. When APR is driven directly from natural-language issue reports, the task remains fundamentally underconstrained: bug reports are often incomplete, many candidate patches appear superficially reasonable, and passing the existing test suite does not guarantee that the reported bug has actually been fixed~\citep{DBLP:conf/issta/QiLAR15,DBLP:journals/corr/abs-2503-15223}. This problem is compounded by the fact that, when an issue is first reported, the existing test suite often does not yet contain a reproduction test for that bug~\cite{DBLP:conf/icse/KangYY23,mundler2024swt, DBLP:journals/corr/abs-2602-10471}.
Bug Reproduction Tests~(BRTs) help close this gap. A BRT is generated from the natural-language issue description and is designed to expose the reported bug. By turning the issue report into an executable, bug-specific signal, BRTs can validate candidate patches, constrain the downstream repair space, and filter out plausible-but-incorrect fixes
~\citep{nashid2025issue2test,DBLP:journals/corr/abs-2603-07326,DBLP:journals/corr/abs-2602-02965}. This role has made BRT generation increasingly important in APR, and has also motivated recent systems to co-generate tests and fixes in iterative loops, where generated tests guide repair and generated fixes in turn shape tests~\citep{DBLP:conf/icml/AhmedGPS0H25,DBLP:journals/corr/abs-2601-19066,li2026beyond}.

\begin{figure*}[t]
    \centering
    \begin{sloppypar}
        \includegraphics[width=0.99\textwidth]{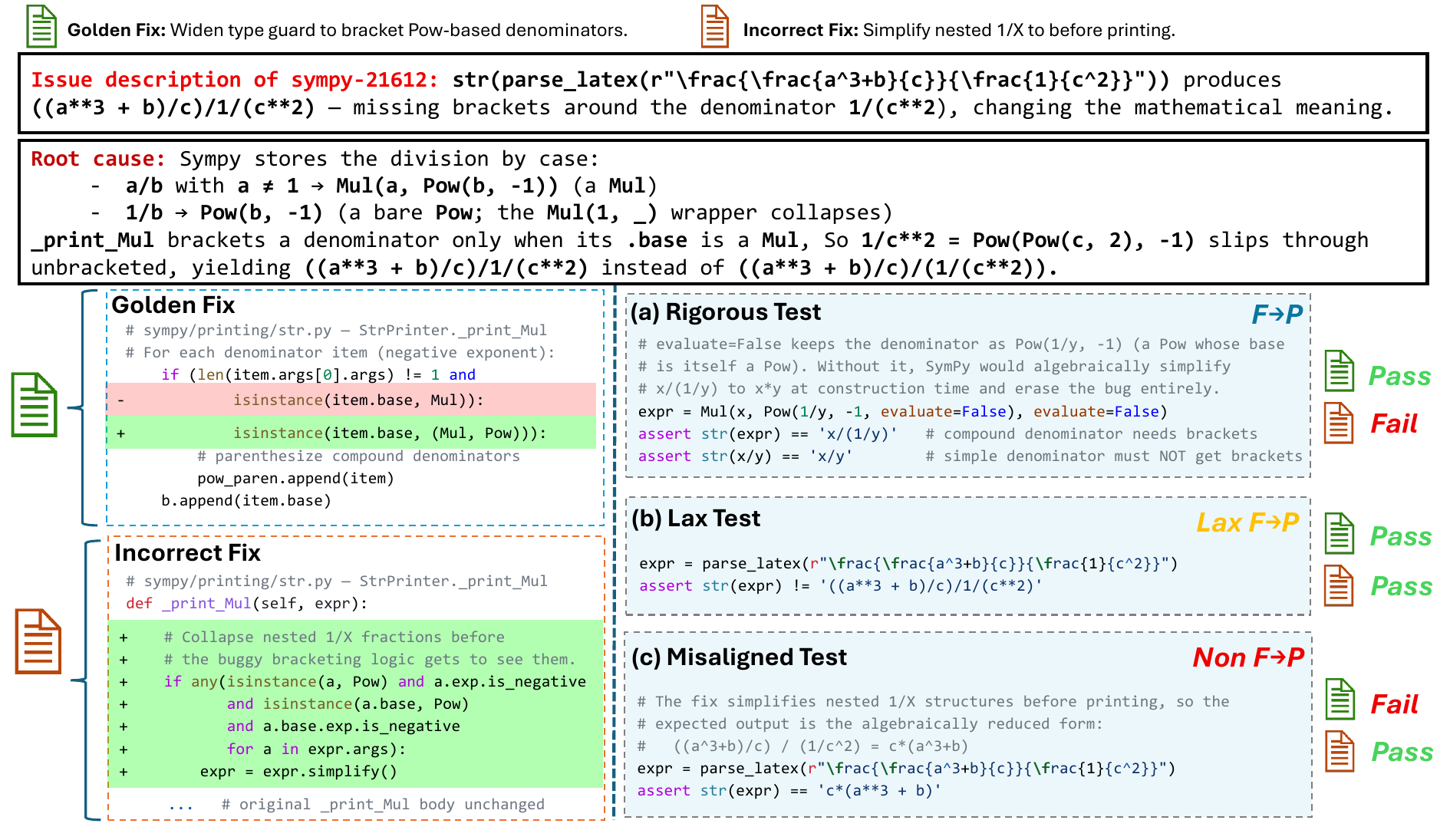}
        \caption{Three reproduction tests for the issue \texttt{sympy-21612} (from SWT-bench~\cite{mundler2024swt}) react differently to two plausible fixes. The bug is: a SymPy printer omits brackets around \texttt{1/X} denominators, changing the math. \textbf{Golden Fix} widens the type check at the root cause. \textbf{Incorrect Fix} instead simplifies away the input pattern which the type check fails to recognize, leaving the check itself untouched. The reproduction tests include: \textbf{(a) Rigorous} test asserts the printer's exact output on two probes, which passes Gold and rejects Incorrect. \textbf{(b) Lax} test only checks the output differs from the known-buggy string, which accepts both fixes indistinguishably. \textbf{(c) Misaligned} test hard-codes the simplified form \texttt{c*(a**3+b)}, which passes Incorrect (which produces this form) and rejects Gold.
        All three tests fail on the buggy code, yet they react differently to the two fixes. Only Rigorous constrains the repair toward the actual root cause.}
        \label{fig:motivation_example}
    \end{sloppypar}
    \vspace{-4mm}
\end{figure*}

The dominant criterion for evaluating generated BRTs is the Fail-to-Pass~(F$\to$P) property: a test must fail on the buggy version and pass on the fixed one. While useful, F$\to$P only checks whether a test detects the reported bug under one reference fix; it does not assess whether the test sufficiently constrains repair by ruling out plausible but incorrect fixes. Figure~\ref{fig:motivation_example} illustrates this with \texttt{sympy-21612}, where a SymPy printer omits brackets around \texttt{1/X} denominators. Three reproduction tests all detect the buggy behavior but differ sharply in repair guidance. A Lax test satisfies F$\to$P yet accepts both the golden fix and an incorrect simplify-away fix. A Rigorous test accepts the golden fix while rejecting the incorrect one. A Misaligned test instead accepts the incorrect fix and rejects the golden fix. Current BRT evaluation does not distinguish these cases: both Rigorous and Lax tests satisfy F$\to$P, although only the Rigorous test properly constrains the repair space.

Current approaches miss this repair-constraining dimension. Methods that generate BRTs separately~\citep{nashid2025issue2test,DBLP:journals/corr/abs-2603-07326} treat the task largely as F$\to$P optimization, while cogeneration methods~\citep{li2026beyond,DBLP:journals/corr/abs-2511-16004} evolve tests and fixes jointly but still do not distinguish Lax from Rigorous tests. Post-hoc variant analysis~\citep{li2026benchmark} similarly uses fix variants to expose inadequate tests, but it requires gold fixes and is aimed at benchmark analysis rather than in-loop repair. As a result, current methods cannot tell whether a generated BRT merely reproduces the symptom or truly constrains repair toward the correct fix.

To study this, we introduce \emph{Mutation Patch Evaluation} (\mpe{}), which evaluates a BRT against plausible-but-incorrect fixes and classifies it as Rigorous, Lax, or Misaligned. Using \mpe{}, we analyze generated BRTs in two settings (\Cref{sec:empirical}): \emph{BRT injection}, where a separately generated BRT guides a fix agent, and \emph{cogeneration}, where one agent jointly produces the fix and BRT. We find that the benefit of F$\to$P-passing BRTs comes entirely from the Rigorous tier ($\Delta\!=\!+8.5$ Resolved), while Lax tests provide no gain ($\Delta\!=\!0.0$). In cogeneration, we further find that the generated test and fix errors are correlated rather than independent, with a joint-failure rate $1.87\times$ the independent prediction.

Building on these findings, we propose \ours{} (\Cref{sec:method}), a two-phase cogeneration loop. \emph{Lax-init} generates a reproduction test $t_0$ before any fix attempt, reducing early Misaligned tests and test--fix coupling. \emph{Hardening} then iteratively refines the $(t_k, c_k)$ pair against a per-round mutant pool, using a \emph{Temporal Matrix} that converts \mpe{}'s within-pool signal into a reference-free in-loop signal. On the SWE-bench~$\cap$~SWT-bench Verified subset, \ours{} reaches $69.4\%$ Resolved and $78.9\%$ \ftp{}, outperforming vanilla cogeneration on the same backbone (OpenHands + GPT-5-mini) by $+8.0$ points in Resolved.
In summary, we make three contributions:
\begin{itemize}[leftmargin=*,itemsep=2pt,topsep=2pt,parsep=0pt]
    \item We introduce \emph{Mutation Patch Evaluation} (\mpe{}), a finer-grained evaluation framework that partitions BRTs into a three-way taxonomy ({Rigorous}, {Lax}, {Misaligned}), exposing a quality dimension orthogonal to F$\to$P.
    \item Using \mpe{}, we conduct an empirical study (\Cref{sec:empirical}) that identifies a research gap between automatic BRT generation and downstream BRT utility. F$\to$P-passing BRTs concentrate their fix-side benefit in the Rigorous tier; cogeneration further produces Lax tests in its own outputs, and its Misaligned tests strongly correlate with incorrect fixes.
    \item We design \ours{}, a two-phase cogeneration loop (lax-init, hardening) that turns \mpe{} into a reference-free in-loop signal via a \emph{Temporal Matrix} (with the previous-round test in place of $t_{\text{golden}}$); the test is hardened against a per-round mutant pool while the agent updates the fix in lockstep.
\end{itemize}

\section{Empirical Study}
\label{sec:empirical}


Bug-reproducing tests (BRTs) are intended to expose a reported bug and verify whether a candidate fix resolves it, but their contribution to bug fixing has not been directly tested. \Cref{fig:motivation_example} shows a failure case where a Lax test satisfies \ftp{} yet still accepts an incorrect fix. We introduce \emph{Mutation Patch Evaluation} (\mpe{}, \Cref{sec:mpe}), which partitions generated BRTs into Rigorous, Lax, and Misaligned, and use it to measure their effect in two settings: a \emph{standalone setting} (\Cref{sec:rq1}), where the BRT is provided to a fix agent as guidance, and a \emph{cogeneration setting}~\citep{DBLP:conf/icml/AhmedGPS0H25, DBLP:journals/corr/abs-2601-19066, li2026beyond, DBLP:journals/corr/abs-2511-16004} (\Cref{sec:rq2}), where one agent jointly produces the fix and the BRT. We find that the \ftp{}-passing pool's fix-side gain concentrates in the Rigorous tier while Lax tests provide none, and that cogeneration's own output exhibits the same Lax pattern coupled with correlated fix errors.

\subsection{Mutation Patch Evaluation (\mpe{})}
\label{sec:mpe}


We follow the standard SWT-Bench setup~\citep{mundler2024swt}, where each instance provides a buggy program $c_{\text{buggy}}$, a golden fix $c_{\text{golden}}$, and a golden test $t_{\text{golden}}$ as the human bug specification. Under this setup, a generated reproduction test $t_{\text{gen}}$ is typically judged by \ftp{}, meaning that it fails on $c_{\text{buggy}}$ and passes on $c_{\text{golden}}$. However, \ftp{} only checks rejection of $c_{\text{buggy}}$ and acceptance of $c_{\text{golden}}$; it does not test whether $t_{\text{gen}}$ rejects plausible but incorrect fixes. Inspired by classical mutation testing~\citep{DBLP:journals/computer/DeMilloLS78, DBLP:journals/tse/JiaH11}, we therefore evaluate $t_{\text{gen}}$ on a pool of \emph{semantic} mutations of $c_{\text{golden}}$ to formalize the three-way classification into Rigorous, Lax, and Misaligned.

\noindent{\bf Mutation patch pool.}
We construct a \emph{mutation patch pool}
$\mathcal{C}=\{c^\prime_1,c^\prime_2,\ldots,c^\prime_N\}$ that approximates the space
of plausible-but-incorrect repairs. 
Since this space is unbounded, we sample it by mutating the
golden fix under a fixed set of \emph{semantic operators}.
Each $c^\prime$ is generated by an LLM under one of five
operators, namely \emph{vanilla}, \emph{symptom suppression},
\emph{incomplete fix}, \emph{input-specific shortcut}, and
\emph{behavior substitution}. These operators target common ways
an incorrect fix can pass a reproduction test. To encourage
mutants in the plausible-but-incorrect region rather than
trivially rejected ones, the mutator is conditioned on
$(c_{\text{golden}}, t_{\text{gen}}, t_{\text{golden}})$ and
targets fixes that pass both tests but leave the bug unresolved.
In \Cref{fig:motivation_example}, the Incorrect Fix simplifies the
input expression, matching the behavior-substitution pattern,
and the Lax test in panel (b) accepts it because the test only
asserts that the output differs from the known-buggy string.
\Cref{app:mpe_pool} gives operator definitions, sampling budget,
and full prompts.

\noindent{\bf Confusion matrix.}
Running each $c^\prime \in \mathcal{C}$ against both $t_{\text{gen}}$
and $t_{\text{golden}}$ yields the following 2$\times$2 confusion matrix.
\begin{center}
\small
\begin{tabular}{c|cc}
\toprule
 & $c^\prime$ \emph{passes} $t_{\text{golden}}$ & $c^\prime$ \emph{fails} $t_{\text{golden}}$ \\
\midrule
$c^\prime$ \emph{passes} $t_{\text{gen}}$ & $\alpha$ (Agree-pass) & $\beta$ (\textbf{Laxity cell}) \\
$c^\prime$ \emph{fails} $t_{\text{gen}}$ & $\gamma$ (\textbf{Rigor cell}) & $\delta$ (Agree-fail) \\
\bottomrule
\end{tabular}
\end{center}


\noindent
The diagonal cells ($\alpha$, $\delta$) collect the mutation patches on
which the generated and golden tests agree, while the two
off-diagonal cells expose failure modes \ftp{} conflates.
Cell~$\beta$ counts mutation patches that pass $t_{\text{gen}}$ but
fail $t_{\text{golden}}$. Such tests detect the bug, yet still
admits incorrect repairs and so under-constrains the repair space.
Cell~$\gamma$ counts patches that fail $t_{\text{gen}}$ but pass
$t_{\text{golden}}$. These tests reject behaviorally valid
repairs and impose inconsistent constraints. 
We define the \emph{Laxity rate} of a test as
$\beta/\mathrm{total}$, the share of mutation patches on which
$t_{\text{gen}}$ is laxer than $t_{\text{golden}}$.


\noindent{\bf Three-way partition.}
Combined with the \ftp{} status of $t_{\text{gen}}$ on
$c_{\text{golden}}$, the 2$\times$2 matrix yields three categories:
Rigorous (\ftp{} with Laxity rate below the Laxity-rate cutoff $\tau$),
where the test detects the bug and constrains repairs consistently with
the golden test; Lax (\ftp{} with Laxity rate at or above $\tau$),
where the test detects the bug but still admits
many plausible yet incorrect repairs; and Misaligned, where the test rejects
$c_{\text{golden}}$ and therefore
conflicts with the golden correctness specification. We use the median
Laxity-rate cutoff $\tau=0.202$ in the main analysis and report cutoff sensitivity in
\Cref{app:cutoff_sensitivity}.


\subsection{Bug Fixing with BRT Injection}
\label{sec:rq1}

We first study how BRTs from \brtonly{} methods affect downstream bug fixing. These methods treat BRT generation as a standalone task and produce a BRT independently of fix generation. We use four representative \brtonly{} methods: LogicStar~\citep{logicstar2025swtbench}, e-Otter++~\citep{ahmed2025heterogeneous}, AssertFlip~\citep{khatib2025assertflip}, and OpenHands in \brtonly{} mode~\citep{DBLP:conf/iclr/0001LSXTZPSLSTL25}.

\noindent{\bf Setup.}
The fix agent is OpenHands with GPT-5-mini. We define \fixonly{} as the same agent given only the issue description with no BRT injected and use it as the no-BRT baseline. For each (instance, BRT) pair from SWT-Bench~\citep{mundler2024swt}, we inject the BRT into the fix agent as test-driven guidance and run the agent once. To measure the BRT's contribution to bug fixing, we use the paired Resolved change on the same instances (the fraction of generated fixes that pass the golden tests). Formally, $\Delta = \mathrm{Resolved}_{\text{with-BRT}} - \mathrm{Resolved}_{\fixonly{}}$. The four \brtonly{} methods together produce $1{,}104$ \ftp{} and $474$ non-\ftp{} pairs (\Cref{app:rq1_setup} shows details of each method).

\noindent\textbf{BRTs that satisfy \ftp{} help the fix agent on average, while those that fail it harm performance.}
On the $1{,}104$ \ftp{} pairs, adding BRTs lifts Resolved from
$68.4\%$ to $72.6\%$ ($\Delta\!=\!+4.3$);
on the $474$ non-\ftp{} pairs, Resolved drops from
$44.5\%$ to $40.9\%$ ($\Delta\!=\!-3.6$).
This validates \ftp{} as a coarse binary filter and rules out the
null hypothesis that all generated BRTs are equally informative.

\noindent\textbf{Within the \ftp{} pool, however, the entire fixing gain concentrates in the Rigorous half.}
We apply \mpe{} (\Cref{sec:mpe}) to split the \ftp{}
pool into Rigorous and Lax halves; \Cref{tab:rq1_tdd}
reports the results.
\begin{table}[tbp]
  \centering
  \small
  \caption{Bug fixing utility of generated BRTs on SWT-Bench.
    $\Delta$ is the paired change in \textsc{Resolved} relative to
    running the same instance without the BRT. The performance gain
    mainly comes from the generated \ftp{} BRTs that are consistent
    with the golden tests (i.e., Rigorous), while the non-\ftp{} BRTs
    (i.e., Misaligned) hurt the fixing performance.}
  \label{tab:rq1_tdd}
  \begin{tabular}{lrrrr}
    \toprule
    Test class & \# instances & \fixonly & with-BRT & $\Delta$ \\
    \midrule
    {Rigorous} & 552 & 67.3\% & 75.8\% & {$+$8.5} \\
    Lax                       & 552 & 69.5\% & 69.5\% & $+$0.0          \\
    Misaligned                & 474 & 44.5\% & 40.9\% & $-$3.6          \\
    \bottomrule
  \end{tabular}
  \vspace{-4mm}
\end{table}

The Rigorous half carries the entire \ftp{}
pool's fixing benefit ($\Delta\!=\!+8.5$), while the
Lax half contributes none ($\Delta\!=\!0.0$).
The same pattern holds within each individual
generator (\Cref{tab:rq1_per_agent} in \Cref{app:rq1_setup}),
ruling out the alternative explanation that the gap is an
artifact of pooling tests across heterogeneous agents.

\begin{rqbox}[title=Finding 1]
\ftp{} agrees with downstream bug fixing utility as a coarse
binary filter. Within the \ftp{} pool, however, the fixing
gain concentrates in the Rigorous half: \ftp{} alone
is too coarse and collapses a within-pool quality dimension
that materially determines BRT utility.
\end{rqbox}

\subsection{Bug Fixing with Cogeneration}
\label{sec:rq2}

In practice, human developers typically resolve a bug by writing the fix and the reproduction test together rather than as two separate tasks. \emph{Cogeneration}~\citep{DBLP:journals/corr/abs-2601-19066} methods mirror this workflow by jointly producing the fix and the BRT in a single trajectory. We study, under this setting, whether the cogenerated BRT actually contributes to downstream bug fixing.

\noindent{\bf Setup.}
We compare three configurations of the same underlying agent
(OpenHands + GPT-5-mini) on SWT-Bench~\citep{mundler2024swt}:
\cogen{} jointly produces a fix and a BRT in one
trajectory~\mbox{\citep{DBLP:journals/corr/abs-2601-19066}}; \brtonly{}
produces only the BRT, evaluated by \ftp{} against the golden fix;
and \fixonly{} produces only the fix, evaluated against the golden
test. Additional setup details are in \Cref{app:rq2_setup}.

\begin{figure}[!t]
\centering
\includegraphics[width=0.75\linewidth]{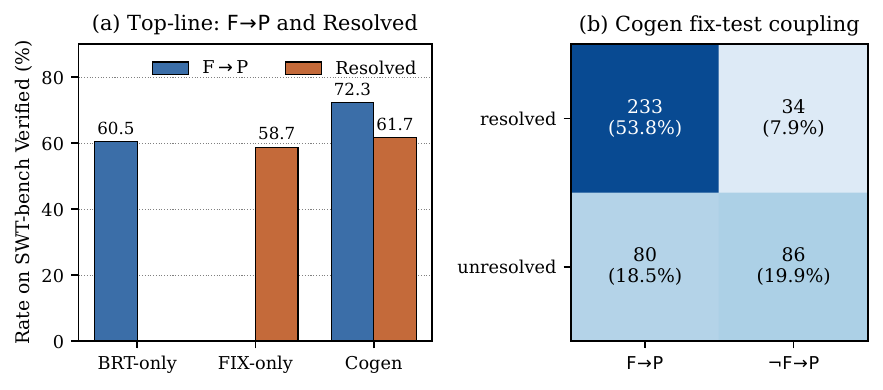}
\caption{%
Cogeneration vs.\ single-task baselines on SWT-Bench.
\textbf{(a)} \cogen{} (joint fix+test in one trajectory) exceeds
each single-task baseline (\brtonly{} on \ftp{}, \fixonly{} on
\textsc{Resolved}), but its fix-side gain is modest.
\textbf{(b)} The Lax pattern of \Cref{sec:rq1} surfaces inside
\cogen{}'s own output. In particular, $80$ of the $166$ unresolved instances
pair an \ftp{}-passing test with an incorrect fix, and the
joint-failure cell is $1.87\times$ the rate predicted under
independent fix and test errors.
}
\label{fig:rq2}
\vspace{-5mm}
\end{figure}

\noindent\textbf{Cogeneration beats both specialists, but the fix-side gain is marginal.}
\Cref{fig:rq2}(a) shows that \cogen{} exceeds each single-task
baseline on both metrics, with test~\ftp{}~$72.3\%$ ($+11.8$ over \brtonly{}) and
\textsc{Resolved}~$61.7\%$ ($+3.0$ over \fixonly{}).

\noindent\textbf{The Lax pattern also appears in the cogeneration setting.}
For $80$ ($48.2\%$) of \cogen{}'s $166$ unresolved instances, the
cogenerated test \ftp{}-passes on the golden fix
(\Cref{fig:rq2}(b), bottom-left cell) yet the cogenerated fix is
wrong. In these trajectories, the agent stopped after observing
its own test pass on its own fix, so the cogenerated test passes
on \emph{both} the wrong fix and the golden fix, exhibiting the Lax
pattern of \Cref{sec:rq1}. A Lax test lets an incorrect fix
satisfy the in-trajectory \ftp{} signal, triggering early stopping
while the bug remains unfixed.

\noindent\textbf{\cogen{}'s fix and test errors are correlated.}
An ideal joint formulation would derive the fix and test from
the issue specification independently, so that errors on each
would be uncorrelated. \Cref{fig:rq2}(b) shows that \cogen{}'s
joint-failure rate is $19.9\%$ ($1.87\times$ the $10.6\%$
predicted under independence), with
$P(\text{fix wrong}\mid\text{test wrong})=71.7\%$ and
$P(\text{test wrong}\mid\text{fix wrong})=51.8\%$. This suggests
that the cogenerated fix and test share a common
(mis)interpretation of the intended behavior.

\begin{rqbox}[title=Finding 2]
Cogeneration's joint fix-side gain over a fix specialist is
marginal. The Lax pattern from \Cref{sec:rq1} resurfaces within
\cogen{}'s own output, and its fix and test errors are correlated
rather than independent because \ftp{} between two artifacts of one
trajectory cannot serve as an independent correctness check.
\end{rqbox}

\section{\ours: Iterative-Hardening Cogeneration}
\label{sec:method}

\begin{figure}[t]
\centering
\includegraphics[width=0.85\linewidth]{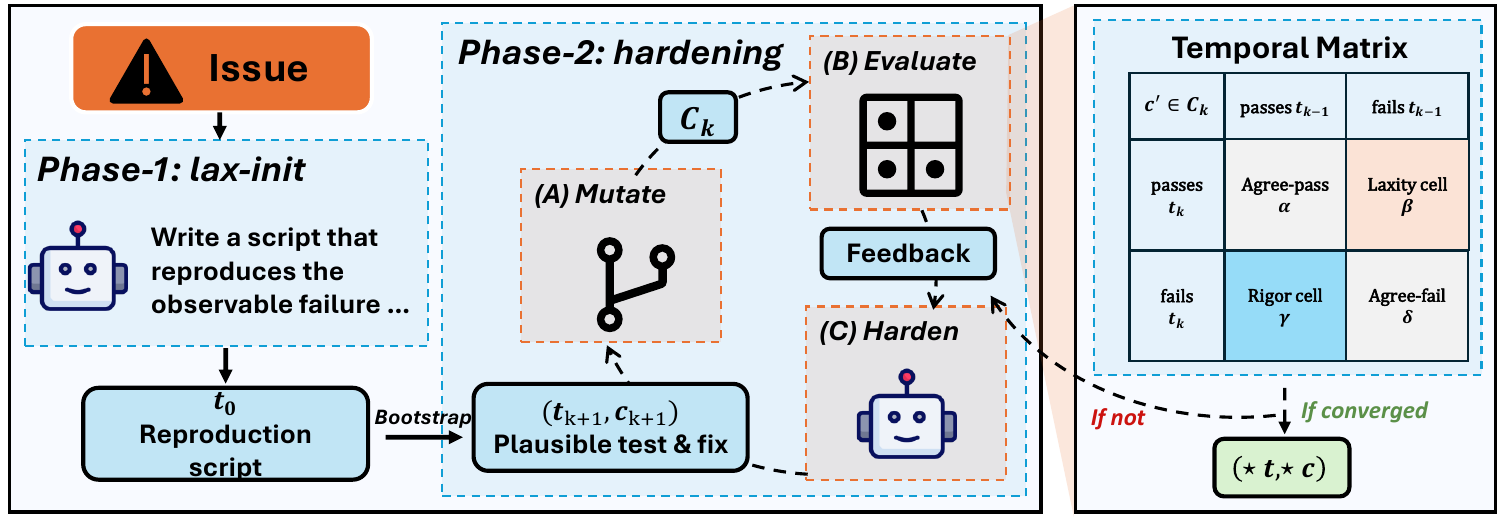}
\caption{%
\textbf{\ours{} pipeline.}
\textbf{Phase~1 (Lax-init):} given the issue, the agent writes a reproduction test $t_0$ targeting the observable failure without source edits.
\textbf{Phase~2 (Hardening):} bootstrapping $t_0$ into a plausible pair $(t_1, c_1)$, the loop iterates three steps per round $k$: \textbf{(A)~Mutate} produces a mutant pool $\mathcal{C}_k$ from $c_k$ via four targeted operators; \textbf{(B)~Evaluate} runs each $c^\prime \in \mathcal{C}_k$ against $t_k$ and $t_{k-1}$ to fill the Temporal Matrix with cells $\alpha$ (Agree-pass), $\beta$ (Laxity), $\gamma$ (Rigor), $\delta$ (Agree-fail); \textbf{(C)~Harden} feeds the Temporal Matrix back to the agent, which strengthens the test (and updates the fix) into the next pair $(t_{k+1}, c_{k+1})$. The loop returns the final pair $(t^\star, c^\star)$ when hardening converges. 
}
\label{fig:method_pipeline}
\end{figure}

We design \ours{} (\Cref{fig:method_pipeline}), a two-phase cogeneration loop, to mitigate two failure modes identified in \Cref{sec:empirical}. Lax tests admit incorrect fixes, while Misaligned tests reject even the golden fix. \emph{Phase~1: Lax-init} (\Cref{sec:method:laxinit}) writes a reproduction test $t_0$ before any fix attempt to avoid an early Misaligned $t_0$. \emph{Phase~2: Hardening} (\Cref{sec:method:hardening}) bootstraps $t_0$ into a $(t_1, c_1)$ pair and iteratively refines it through rounds of \emph{Mutate}, \emph{Evaluate}, and \emph{Harden}, with a Temporal Matrix flagging per-round Lax regression. We detail Phase~1 in \Cref{sec:method:laxinit} and Phase~2 in \Cref{sec:method:hardening}.

\subsection{Lax-init: produce the test before the fix}
\label{sec:method:laxinit}

Misaligned tests harm the fix process. \Cref{sec:rq1} shows that
injecting a non-\ftp{} (Misaligned) BRT into the fix agent lowers
Resolved below the no-BRT baseline: the wrong test
misleads the agent rather than helping it. The same coupling
appears inside cogeneration, where the cogenerated fix tends to
be wrong when the cogenerated test is wrong (\Cref{sec:rq2}).

To avoid Misaligned $t_0$, \ours{} mirrors how human developers
begin debugging: write a short script that captures the
observable failure (printed output, raised exception,
return-value discrepancy) before reasoning about the
implementation. The agent therefore emits only a reproduction
test $t_0$ and is forbidden from modifying any source file; the
prompt steers $t_0$ toward the most observable witness of the
bug and away from internal hooks or strict structural assertions.
Writing the test first also weakens the cogeneration coupling of
\Cref{sec:rq2}, since $t_0$ has no current fix to anchor against.


\subsection{Hardening: cogenerate and tighten the (test, fix) pair}
\label{sec:method:hardening}

After Phase~1 produces $t_0$, Hardening iteratively refines a
$(t_k, c_k)$ pair over rounds $k=1,\ldots,K$. At each round, the
agent debugs from the previous round to locate the root cause
and jointly emit an \ftp{}-passing pair $(t_k, c_k)$, where $t_k$
fails on the buggy code with $c_k$ reverted, and passes on
$c_k$. Joint generation gives the test concrete internal
behaviors to check, and \Cref{sec:rq2} provides empirical
support that cogeneration raises test \ftp{} above the BRT-only
baseline.

The \ftp{} gate is binary, however, and cannot tell Rigorous
from Lax, so passing it is not by itself a sign of progress.
In \Cref{sec:mpe}, $c_{\text{golden}}$ seeds the mutant pool,
while $t_{\text{golden}}$ serves as the reference test for grading
$t_{\text{gen}}$. Inside the loop, neither golden artifact is
available. \ours{} therefore generates mutants from the current fix
$c_k$ and grades the current test $t_k$ against the previous-round
test $t_{k-1}$.
\label{sec:method:signal}
At each round $k$, after the agent emits an \ftp{}-passing
$(t_k, c_k)$, the loop applies the four targeted operators of
\Cref{sec:mpe} to $c_k$ to produce a mutant pool
$\mathcal{C}_k$, and running each $c^\prime \in \mathcal{C}_k$
against both $t_k$ and $t_{k-1}$ yields the \emph{Temporal
Matrix}:

\begin{center}
\small
\begin{tabular}{c|cc}
\toprule
 & $c^\prime$ \emph{passes} $t_{k-1}$ & $c^\prime$ \emph{fails} $t_{k-1}$ \\
\midrule
$c^\prime$ \emph{passes} $t_k$ & $\alpha$ (Agree-pass) & $\beta$ (\textbf{Laxity cell}) \\
$c^\prime$ \emph{fails} $t_k$ & $\gamma$ (\textbf{Rigor cell}) & $\delta$ (Agree-fail) \\
\bottomrule
\end{tabular}
\end{center}



\noindent
The resulting Temporal Matrix measures whether the new test $t_k$
moves from Lax toward Rigorous relative to $t_{k-1}$. As the empirical
study in \Cref{sec:rq1} shows, $\beta/\mathrm{total}$ separates Lax
tests from Rigorous ones, where $\beta$ counts mutants accepted by the
generated test but rejected by the reference test. \ours{} therefore
uses $\ell_k=\beta/\mathrm{total}$ as its in-loop Laxity rate, where
$\mathrm{total}$ is the number of admissible mutants in $\mathcal{C}_k$.
We set the convergence rule to accept a round only when the updated
test is non-trivially changed and its Laxity rate stays below
$\tau=0.2$, matching the empirical Rigorous/Lax boundary in
\Cref{sec:rq1}. If $\ell_k>\tau$, the update remains too Lax and the
loop continues with feedback from the surviving mutants. If $t_k$
changes non-trivially and $\ell_k\le\tau$, the round is clean and the
loop exits. \Cref{app:method_impl} gives the full algorithm.

\section{Experiments} \label{sect:exp}



\noindent\textbf{Benchmark.}
The main experiments use the 433-instance intersection of SWE-bench Verified~\citep{DBLP:conf/iclr/JimenezYWYPPN24} and SWT-bench Verified~\citep{mundler2024swt}, enabling joint evaluation of fix correctness and test quality on the same instances. For cross-benchmark generalization (\Cref{sec:rq2_general}), we additionally report results on SWE-bench Lite.

\noindent\textbf{Metrics.}
\textbf{Resolved} is the fraction of instances on which the generated fix passes the golden test (SWE-bench harness). \textbf{\ftp{} rate} is the fraction of instances on which the generated test fails on the buggy code and passes on the golden fix (SWT-bench harness).

\noindent\textbf{Baselines.}
We compare \ours{} against two families of methods: \textbf{(i)~fix-only} agents that generate only a fix, and \textbf{(ii)~cogeneration} methods that jointly produce the fix and the BRT in a single trajectory. \Cref{tab:main_fix} lists all baselines; detailed descriptions are in \Cref{app:baselines}.

\noindent\textbf{Implementation.}
We instantiate \ours{} on top of OpenHands with GPT-5-mini as the backbone LLM, the same backbone used in \Cref{sec:empirical}. The hardening loop runs for $K=5$ rounds; each round generates the mutant pool $\mathcal{C}_k$ with the four targeted operators of \Cref{sec:mpe} ($N=3$ samples per operator, $|\mathcal{C}_k|=12$ at full admission). All OpenHands-based agents (\ours{}, OpenHands fix-only, OpenHands\,+\,\cogen{}) use GPT-5-mini for fair comparison. For other agents, we report scores from each method's official release. Per-round implementation details and the pluggable critic interface are in \Cref{app:method_impl,app:skill_critic}.

\subsection{\ours{} Improves Fix Correctness over Existing Pipelines}
\label{sec:rq1_main}



\Cref{tab:main_fix} compares \ours{} against fix-only and cogeneration baselines.

\noindent\textbf{\ours{} outperforms all baselines on Resolved.}
\ours{} achieves $69.4\%$ Resolved, the highest in \Cref{tab:main_fix}. Building on \Cref{sec:rq2}, \ours{} introduces a two-phase design (lax-init + hardening) that directly addresses limitations of vanilla cogeneration, yielding a $+8.0$ gain in Resolved over the same-backbone OpenHands+\cogen{} baseline. Other cogeneration baselines (InfCode $61.5\%$, Agent-CoEvo $58.2\%$) and the strongest fix-only agent (SWE-Agent $59.8\%$) sit below this mark.

\noindent\textbf{\ours{} improves both test effectiveness and fix correctness.}
\ours{}'s \ftp{} rate ($78.9\%$) exceeds OpenHands+\cogen{}'s ($72.7\%$) by $+6.2$ points, while its Resolved rate improves by $+8.0$ points. Against InfCode, \ours{} gains $+9.3$ points in \ftp{} and $+7.9$ points in Resolved. This joint improvement is consistent with the empirical finding of \Cref{sec:empirical}: within the \ftp{} pool, fixing utility concentrates in the Rigorous tier. The iterative hardening loop (\Cref{sec:method:hardening}) is designed to shift tests from Lax to Rigorous by killing surviving mutants round over round, so stronger reproduction tests translate into stronger downstream fixes.

\noindent\textbf{Cost remains practical.}
\ours{} costs \$0.84 per instance, comparable to existing cogeneration methods (InfCode \$0.77, Agent-CoEvo \$0.88). \Cref{sec:rq3_ablation} further shows that most of the Resolved gain is obtained within the first few rounds; \Cref{tab:app_round_cost} reports the per-round active-instance counts and shows that MPE mutant generation accounts for 9.5\% of the final run cost.

\begin{table}[t]
  \centering
  \small
  \caption{Resolved, \ftp{}, and average per-instance cost on the SWE-bench Verified $\cap$ SWT-bench Verified setting. All methods share the same GPT-5-mini backbone. ``\textendash'' denotes unreported values. \ours{} outperforms fix-only and cogeneration baselines on both Resolved and \ftp{}, suggesting that iterative hardening improves fix quality while producing stronger reproduction tests.}
  \label{tab:main_fix}
  \begin{tabular}{lccc}
    \toprule
    Method & Resolved (\%) & Test F$\to$P (\%) & Avg.~cost (USD/inst) \\
    \midrule
    \multicolumn{4}{l}{\emph{\fixonly{} (no reproduction test)}} \\
    \midrule
    SWE-Agent~\citep{DBLP:conf/nips/YangJWLYNP24}               & 59.8         & --            & 0.10        \\
  mini SWE-Agent~\citep{DBLP:conf/nips/YangJWLYNP24}                                & 55.3     & --            & 0.17  \\
      Agentless~\citep{DBLP:journals/pacmse/XiaDDZ25}                          & 46.5         & --            & 0.24      \\
      OpenHands~\citep{DBLP:conf/iclr/0001LSXTZPSLSTL25}          & 58.9          & --            & 0.39      \\
    \midrule
    \multicolumn{4}{l}{Cogeneration \emph{(fix and BRT produced jointly)}} \\
    \midrule
      InfCode~\citep{DBLP:journals/corr/abs-2511-16004}           & 61.5         & 69.6      & 0.77      \\
      Agent-CoEvo~\citep{li2026beyond}                                  & 58.2      & 64.9      & 0.88  \\
      OpenHands\,+\,\cogen~\citep{DBLP:conf/iclr/0001LSXTZPSLSTL25}
                                                                  & 61.4          & 72.7          & 0.56      \\
      \textbf{\ours{}} (Ours)                                     & \textbf{69.4} & \textbf{78.9} & 0.84  \\
    \bottomrule
  \end{tabular}
\end{table}

\subsection{\ours{} Generalizes across LLM Backbones and Benchmarks}
\label{sec:rq2_general}

\begin{table}[t]
  \centering
  \small
  \caption{Cogeneration methods across LLM backbones (GPT-5.4 and Claude Opus~4.5) on the 433-instance SWE-bench Verified $\cap$ SWT-bench Verified: InfCode~\citep{DBLP:journals/corr/abs-2511-16004}, Agent-CoEvo~\citep{li2026beyond}, and OpenHands\,+\,\cogen{}~\citep{DBLP:conf/iclr/0001LSXTZPSLSTL25}. \ours{} leads under both backbones, with up to $+4.0$ Resolved gain over the strongest baseline.}
  \label{tab:cogen_models}
  \begin{tabular}{lcccccc}
    \toprule
    & \multicolumn{2}{c}{Resolved (\%)} & \multicolumn{2}{c}{Test F$\to$P (\%)} & \multicolumn{2}{c}{Avg.~cost (USD/inst)} \\
    \cmidrule(lr){2-3} \cmidrule(lr){4-5} \cmidrule(lr){6-7}
    Method & GPT-5.4 & Opus~4.5 & GPT-5.4 & Opus~4.5 & GPT-5.4 & Opus~4.5 \\
    \midrule
    InfCode                  & 73.2 & 74.4 & 76.6 & 71.3 &  7.25 & 26.91 \\
    Agent-CoEvo              & 73.5 & 74.7 & 75.6 & 71.8 &  8.79 & 28.43 \\
    OpenHands\,+\,\cogen     & 70.0 & 70.8 & 75.3 & 71.5 &  4.26 & 13.99 \\
    \textbf{\ours{}} (Ours)  & \textbf{76.2} & \textbf{78.7} & \textbf{76.7} & \textbf{79.0} & 6.69 & 18.07 \\
    \bottomrule
  \end{tabular}
\end{table}

\noindent\textbf{Across LLM backbones.}
To test whether \ours{}'s gains are tied to a specific backbone, we re-run cogeneration methods with GPT-5.4 and Claude Opus~4.5 on the 433-instance SWE-bench Verified $\cap$ SWT-bench Verified (\Cref{tab:cogen_models}). \ours{} achieves $76.2\%$ Resolved with GPT-5.4 and $78.7\%$ with Claude Opus~4.5, outperforming the strongest cogeneration baseline (Agent-CoEvo, $73.5\%$ / $74.7\%$) by $+2.7$ and $+4.0$ points respectively. \ours{}'s per-instance cost (\$6.69 / \$18.07) also sits below InfCode (\$7.25 / \$26.91) and Agent-CoEvo (\$8.79 / \$28.43) under both backbones. The consistent improvement across two architecturally different LLMs indicates that the hardening mechanism is backbone-agnostic: the gain comes from the loop structure rather than from a model-specific ability.

\noindent\textbf{Across benchmarks.}
\Cref{tab:cogen_swebench_lite} reports results on the 276-instance SWE-bench Lite $\cap$ SWT-bench Lite. \ours{} reaches $57.6\%$ Resolved, outperforming InfCode ($54.2\%$) by $+3.4$ points and OpenHands+\cogen{} ($52.0\%$) by $+5.6$ points. The consistent improvement on a separate benchmark confirms that \ours{} is not overfit to the SWE-bench Verified task distribution.

\begin{table}[t]
  \centering
  \small
\caption{Cogeneration methods on the 276-instance SWE-bench Lite $\cap$ SWT-bench Lite with GPT-5-mini. \ours{} achieves the highest Resolved ($57.6\%$) and \ftp{} ($64.1\%$), confirming that \ours{} transfers to a different benchmark.}
  \label{tab:cogen_swebench_lite}
  \begin{tabular}{lcc}
    \toprule
    Method & Resolved (\%) & Test F$\to$P (\%) \\
    \midrule
    OpenHands\,+\,\cogen~\citep{DBLP:journals/corr/abs-2601-19066}
            & 52.0 & 62.3 \\
    Agent-CoEvo~\citep{li2026beyond}
            & 51.0 & 53.6 \\
    InfCode~\citep{DBLP:journals/corr/abs-2511-16004}
            & 54.2 & 63.7 \\
    \textbf{\ours{}} (Ours)
            & \textbf{57.6} & \textbf{64.1} \\
    \bottomrule
  \end{tabular}
\end{table}


\subsection{Ablation Study: Effects of Lax-init and Hardening}
\label{sec:rq3_ablation}

\begin{figure}[t]
\centering
\begin{minipage}[c]{0.46\textwidth}
\centering
\small
\captionof{table}{Phase ablation. Each variant removes one
phase from the full \ours{} pipeline. Lax-init and hardening
each address a distinct cogeneration weakness; either alone
recovers little of the full gain.}
\label{tab:rq3_ablation}
\begin{tabular}{lcc}
  \toprule
  Variant & Resolved (\%) & Test F$\to$P (\%) \\
  \midrule
  \textbf{\ours{}}              & \textbf{69.4} & 78.9 \\
  \quad w/o lax-init            & 62.4          & 70.2 \\
  \quad w/o hardening           & 61.0          & \textbf{82.6} \\
  \quad w/o both                & 61.4          & 72.7 \\
  \bottomrule
\end{tabular}
\end{minipage}\hfill
\begin{minipage}[c]{0.5\textwidth}
\centering
\setlength{\abovecaptionskip}{2pt}
\includegraphics[width=\linewidth]{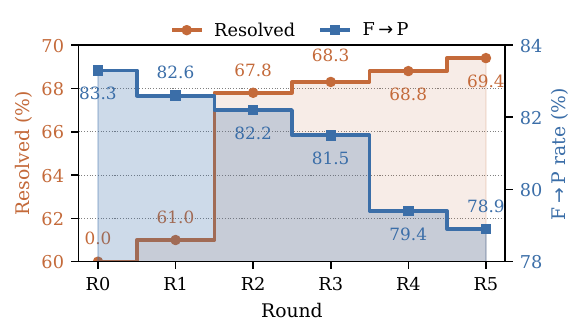}
\captionof{figure}{Per-round evaluation of Resolved and \ftp{}.
Resolved rises sharply after the first hardening step and then improves gradually; \ftp{} decreases from $83.3\%$ at R0 to $78.9\%$.} 
\label{fig:step_dynamics}
\end{minipage}
\end{figure}

\ours{} has two design modules aligned with the weaknesses in \Cref{sec:empirical}. Lax-init targets early test--fix misalignment, while hardening targets Lax acceptance of incorrect fixes. We assess both modules with the phase ablation in \Cref{tab:rq3_ablation} and the per-round dynamics in \Cref{fig:step_dynamics}.

\smallskip
\textbf{Lax-init reduces early misalignment.}
By generating the test before any fix is committed, lax-init reduces the chance that the initial $(t, c)$ pair shares the same wrong interpretation. \Cref{tab:rq3_ablation} supports this: removing lax-init drops Resolved to $62.4\%$ ($\Delta\!=\!{-7.0}$); even with the hardening loop still applied, the result is only $+1.0$ over vanilla \cogen{} ($61.4\%$). This indicates that once the initial $(t, c)$ pair is jointly misaligned, hardening alone cannot fully recover.

\smallskip
\textbf{Hardening discourages Lax acceptance, with saturation after R3.}
Hardening iteratively strengthens tests and filters regressions (\Cref{sec:method:hardening}), reducing Lax acceptance. Without it, Resolved drops to $61.0\%$ ($\Delta\!=\!{-8.4}$) even though \ftp{} remains high at $82.6\%$, a typical Lax signature (\Cref{tab:rq3_ablation}). A budget-matched Ralph self-refinement baseline ($K=5$) reaches $63.8\%$ Resolved and $66.7\%$ \ftp{} (Appendix~\Cref{tab:app_ralph_baseline}), showing that extra refinement rounds alone do not explain \ours{}'s joint gains. Across hardening rounds, Resolved jumps from $61.0\%$ (R1) to $67.8\%$ (R2), then improves more gradually to $69.4\%$ in the final output; \ftp{} decreases from $83.3\%$ at R0 to $78.9\%$ at R5 as the test is tightened. Most gains occur within the first few rounds; later improvements are marginal.


\section{Related Work}

\subsection{Generating and Evaluating Bug Reproduction Tests}

Prior work on bug reproduction tests (BRTs) focuses on generating tests from issue descriptions and judging them by F$\to$P. Methods improve generation through issue-driven or repository-aware construction (Issue2Test~\citep{nashid2025issue2test}, Otter~\citep{DBLP:conf/icml/AhmedGPS0H25}), retrieval and execution feedback (Echo~\citep{DBLP:journals/corr/abs-2603-07326}), model specialization (SWE-tester~\citep{soni2026swe}), or test inversion from passing tests (AssertFlip~\citep{khatib2025assertflip}). Related studies ask whether such tests improve downstream repair~\citep{DBLP:journals/ese/ZhangWXLLL26} or analyze empirical properties of generated tests~\citep{DBLP:journals/corr/abs-2602-02965}. All evaluate test quality through F$\to$P or repair utility rather than whether a test constrains plausible fixes; \ours{} fills this gap by turning the Rigorous/Lax/Misaligned categories of \Cref{sec:rq1} into an in-loop generation signal.


\subsection{Test-Fix Co-Evolution}

Another line of work treats tests and fixes as coupled search objects and improves them jointly. Dynamic Cogeneration~\citep{DBLP:journals/corr/abs-2601-19066} studies different loop orderings, while Agent-CoEvo~\citep{li2026beyond} and InfCode~\citep{DBLP:journals/corr/abs-2511-16004} iteratively refine tests and fixes through mutual or adversarial feedback. CoCoEvo~\citep{li2025cocoevo} and LLMLOOP~\citep{ravi2025llmloop} explore similar co-evolutionary designs in function-level settings. Together, these works show that tests should not be generated independently of fixes. However, their interaction signals remain undifferentiated because the loop does not distinguish beneficial rigor from laxity. Our work targets this gap by identifying Lax and Misaligned tests and addressing the resulting test-fix coupling and repair failures.


\subsection{Overfitting and Variant-Based Test Analysis}

The broader repair literature has long shown that passing available tests is not enough for correctness. Work on patch plausibility~\citep{DBLP:conf/issta/QiLAR15} and behavioral divergence in SWE-bench-style evaluation~\citep{DBLP:journals/corr/abs-2503-15223} highlights the limits of weak tests and motivates evaluating how well tests constrain variants. SWE-ABS~\citep{DBLP:journals/corr/abs-2603-00520} uses adversarial analysis to expose inflated success rates under weak tests, SWE-Bench+~\citep{DBLP:journals/corr/abs-2410-06992} reveals benchmark blind spots, EVOREPAIR~\citep{DBLP:conf/kbse/Soto19} studies co-evolution for patch quality, and STING~\citep{li2026benchmark} uses variant pools to expose under-constrained behavior. However, these approaches are mainly benchmark-oriented, post-hoc, or dependent on golden artifacts, and they do not provide a repair-oriented taxonomy that separates Lax from Misaligned tests and assigns distinct corrective actions.

\section{Discussion}
\label{sec:discussion}

\smallskip\noindent\textbf{Beyond \ftp{}.} \ftp{} is a useful but coarse criterion for BRT generation; within the \ftp{}-passing pool the entire fix-side gain concentrates in the Rigorous tier (\Cref{sec:rq1}). \mpe{} surfaces this within-pool dimension as a reference-free signal usable both at evaluation time and inside the cogeneration loop, suggesting that test-quality metrics for repair should look beyond binary specification matching.

\smallskip\noindent\textbf{Why hardening saturates.} Resolved plateaus after R3 (\Cref{sec:rq3_ablation}). One plausible factor is that hardening still cogenerates the test and the fix in lockstep, so the test--fix coupling of \Cref{sec:rq2} can re-emerge across rounds: the test tightens around what the current fix already passes and the fix updates around what the current test already accepts, leaving certain joint errors invisible to the Temporal Matrix. Breaking this in-loop coupling is a promising direction for future work.

\smallskip\noindent\textbf{Limitations.} \mpe{} and the Temporal Matrix rely on LLM-synthesized mutants of $c_{\text{golden}}$ or $c_k$, so their coverage inherits the underlying LLM's notion of plausible-but-incorrect fixes; the four targeted operators are tuned to common Lax patterns we observed (\Cref{sec:mpe}), and rare or domain-specific Lax patterns may go undetected. Our experiments focus on Python bug-fix benchmarks (SWE-bench Verified and SWE-bench Lite) and OpenHands-style agent loops. Generalization to other languages, repair settings such as security patching, or non-agentic LLM systems remains untested.

\section{Conclusion}
\label{sec:conclusion}

We studied whether generated bug-reproducing tests (BRTs) actually advance downstream bug fixing. To answer this, we introduced \emph{Mutation Patch Evaluation} (\mpe{}), which classifies BRTs as Rigorous, Lax, or Misaligned. Our empirical study revealed two limitations of \ftp{}-only quality control: the fix-side gain concentrates in the Rigorous tier, while cogeneration produces Lax tests with correlated fix errors. Building on these findings, we proposed \ours{}, a two-phase cogeneration loop that uses lax-init to avoid early misalignment and hardening with a Temporal Matrix to discourage Lax acceptance. \ours{} reaches $69.4\%$ Resolved and $78.9\%$ \ftp{} on SWE-bench Verified, outperforming the strongest fix-only and cogeneration baselines by $+9.6$ and $+7.9$ percentage points in Resolved, respectively, with consistent gains across LLM backbones and benchmarks.


\newpage
\bibliography{ref}

@inproceedings{DBLP:conf/icse/XiaWZ23,
  author       = {Chunqiu Steven Xia and
                  Yuxiang Wei and
                  Lingming Zhang},
  title        = {Automated Program Repair in the Era of Large Pre-trained Language
                  Models},
  booktitle    = {45th {IEEE/ACM} International Conference on Software Engineering,
                  {ICSE} 2023, Melbourne, Australia, May 14-20, 2023},
  pages        = {1482--1494},
  publisher    = {{IEEE}},
  year         = {2023},
  url          = {https://doi.org/10.1109/ICSE48619.2023.00129},
  doi          = {10.1109/ICSE48619.2023.00129},
  bibsource    = {dblp computer science bibliography, https://dblp.org}
}

@inproceedings{DBLP:conf/nips/YangJWLYNP24,
  author       = {John Yang and
                  Carlos E. Jimenez and
                  Alexander Wettig and
                  Kilian Lieret and
                  Shunyu Yao and
                  Karthik Narasimhan and
                  Ofir Press},
  editor       = {Amir Globersons and
                  Lester Mackey and
                  Danielle Belgrave and
                  Angela Fan and
                  Ulrich Paquet and
                  Jakub M. Tomczak and
                  Cheng Zhang},
  title        = {SWE-agent: Agent-Computer Interfaces Enable Automated Software Engineering},
  booktitle    = {Advances in Neural Information Processing Systems 38: Annual Conference
                  on Neural Information Processing Systems 2024, NeurIPS 2024, Vancouver,
                  BC, Canada, December 10 - 15, 2024},
  year         = {2024},
  url          = {http://papers.nips.cc/paper\_files/paper/2024/hash/5a7c947568c1b1328ccc5230172e1e7c-Abstract-Conference.html},
  bibsource    = {dblp computer science bibliography, https://dblp.org}
}

@inproceedings{DBLP:conf/issta/0002RFR24,
  author       = {Yuntong Zhang and
                  Haifeng Ruan and
                  Zhiyu Fan and
                  Abhik Roychoudhury},
  editor       = {Maria Christakis and
                  Michael Pradel},
  title        = {AutoCodeRover: Autonomous Program Improvement},
  booktitle    = {Proceedings of the 33rd {ACM} {SIGSOFT} International Symposium on
                  Software Testing and Analysis, {ISSTA} 2024, Vienna, Austria, September
                  16-20, 2024},
  pages        = {1592--1604},
  publisher    = {{ACM}},
  year         = {2024},
  url          = {https://doi.org/10.1145/3650212.3680384},
  doi          = {10.1145/3650212.3680384},
  bibsource    = {dblp computer science bibliography, https://dblp.org}
}

@article{DBLP:journals/pacmse/XiaDDZ25,
  author       = {Chunqiu Steven Xia and
                  Yinlin Deng and
                  Soren Dunn and
                  Lingming Zhang},
  title        = {Demystifying LLM-Based Software Engineering Agents},
  journal      = {Proc. {ACM} Softw. Eng.},
  volume       = {2},
  number       = {{FSE}},
  pages        = {801--824},
  year         = {2025},
  url          = {https://doi.org/10.1145/3715754},
  doi          = {10.1145/3715754},
  bibsource    = {dblp computer science bibliography, https://dblp.org}
}

@article{nashid2025issue2test,
  author       = {Noor Nashid and
                  Islem Bouzenia and
                  Michael Pradel and
                  Ali Mesbah},
  title        = {Issue2Test: Generating Reproducing Test Cases from Issue Reports},
  journal      = {CoRR},
  volume       = {abs/2503.16320},
  year         = {2025},
  url          = {https://doi.org/10.48550/arXiv.2503.16320},
  doi          = {10.48550/ARXIV.2503.16320},
  eprinttype   = {arXiv},
  eprint       = {2503.16320},
  bibsource    = {dblp computer science bibliography, https://dblp.org}
}

@article{DBLP:journals/ese/ZhangWXLLL26,
  author       = {Chengming Zhang and
                  Haoye Wang and
                  Chuyang Xu and
                  Jiakun Liu and
                  Kui Liu and
                  Zhongxin Liu},
  title        = {Can test cases generated by large language models facilitate automated
                  program repair?},
  journal      = {Empir. Softw. Eng.},
  volume       = {31},
  number       = {3},
  pages        = {68},
  year         = {2026},
  url          = {https://doi.org/10.1007/s10664-026-10802-w},
  doi          = {10.1007/S10664-026-10802-W},
  bibsource    = {dblp computer science bibliography, https://dblp.org}
}

@article{DBLP:journals/corr/abs-2603-07326,
  author       = {Zhiwei Fei and
                  Yue Pan and
                  Federica Sarro and
                  Jidong Ge and
                  Marc Liu and
                  Vincent Ng and
                  He Ye},
  title        = {Echo: Graph-Enhanced Retrieval and Execution Feedback for Issue Reproduction
                  Test Generation},
  journal      = {CoRR},
  volume       = {abs/2603.07326},
  year         = {2026},
  url          = {https://doi.org/10.48550/arXiv.2603.07326},
  doi          = {10.48550/ARXIV.2603.07326},
  eprinttype   = {arXiv},
  eprint       = {2603.07326},
  bibsource    = {dblp computer science bibliography, https://dblp.org}
}

@article{DBLP:journals/corr/abs-2602-02965,
  author       = {Andr{\'{e}} C. Hora and
                  Gordon Fraser},
  title        = {Understanding Bug-Reproducing Tests: {A} First Empirical Study},
  journal      = {CoRR},
  volume       = {abs/2602.02965},
  year         = {2026},
  url          = {https://doi.org/10.48550/arXiv.2602.02965},
  doi          = {10.48550/ARXIV.2602.02965},
  eprinttype   = {arXiv},
  eprint       = {2602.02965},
  bibsource    = {dblp computer science bibliography, https://dblp.org}
}

@article{DBLP:journals/corr/abs-2601-19066,
  author       = {Runxiang Cheng and
                  Michele Tufano and
                  Jos{\'{e}} Cambronero and
                  Renyao Wei and
                  Sherry Shi and
                  Grant Uy and
                  Pat Rondon and
                  Franjo Ivancic},
  title        = {Dynamic Cogeneration of Bug Reproduction Test in Agentic Program Repair},
  journal      = {CoRR},
  volume       = {abs/2601.19066},
  year         = {2026},
  url          = {https://doi.org/10.48550/arXiv.2601.19066},
  doi          = {10.48550/ARXIV.2601.19066},
  eprinttype   = {arXiv},
  eprint       = {2601.19066},
  bibsource    = {dblp computer science bibliography, https://dblp.org}
}

@article{li2026beyond,
  title={Beyond Fixed Tests: Repository-Level Issue Resolution as Coevolution of Code and Behavioral Constraints},
  author={Li, Kefan and Yuan, Yuan and Wang, Mengfei and Zheng, Shihao and Wang, Wei and Yang, Ping and Li, Mu and Lv, Weifeng},
  journal={arXiv preprint arXiv:2604.04580},
  year={2026}
}

@inproceedings{DBLP:conf/icml/AhmedGPS0H25,
  author       = {Toufique Ahmed and
                  Jatin Ganhotra and
                  Rangeet Pan and
                  Avraham Shinnar and
                  Saurabh Sinha and
                  Martin Hirzel},
  editor       = {Aarti Singh and
                  Maryam Fazel and
                  Daniel Hsu and
                  Simon Lacoste{-}Julien and
                  Felix Berkenkamp and
                  Tegan Maharaj and
                  Kiri Wagstaff and
                  Jerry Zhu},
  title        = {Otter: Generating Tests from Issues to Validate {SWE} Patches},
  booktitle    = {Forty-second International Conference on Machine Learning, {ICML}
                  2025, Vancouver, BC, Canada, July 13-19, 2025},
  series       = {Proceedings of Machine Learning Research},
  publisher    = {{PMLR} / OpenReview.net},
  year         = {2025},
  url          = {https://proceedings.mlr.press/v267/ahmed25b.html},
  bibsource    = {dblp computer science bibliography, https://dblp.org}
}

@inproceedings{DBLP:conf/issta/QiLAR15,
  author       = {Zichao Qi and
                  Fan Long and
                  Sara Achour and
                  Martin C. Rinard},
  editor       = {Michal Young and
                  Tao Xie},
  title        = {An analysis of patch plausibility and correctness for generate-and-validate
                  patch generation systems},
  booktitle    = {Proceedings of the 2015 International Symposium on Software Testing
                  and Analysis, {ISSTA} 2015, Baltimore, MD, USA, July 12-17, 2015},
  pages        = {24--36},
  publisher    = {{ACM}},
  year         = {2015},
  url          = {https://doi.org/10.1145/2771783.2771791},
  doi          = {10.1145/2771783.2771791},
  bibsource    = {dblp computer science bibliography, https://dblp.org}
}

@article{DBLP:journals/corr/abs-2503-15223,
  author       = {You Wang and
                  Michael Pradel and
                  Zhongxin Liu},
  title        = {Are "Solved Issues" in SWE-bench Really Solved Correctly?
                  An Empirical Study},
  journal      = {CoRR},
  volume       = {abs/2503.15223},
  year         = {2025},
  url          = {https://doi.org/10.48550/arXiv.2503.15223},
  doi          = {10.48550/ARXIV.2503.15223},
  eprinttype   = {arXiv},
  eprint       = {2503.15223},
  bibsource    = {dblp computer science bibliography, https://dblp.org}
}

@article{li2026benchmark,
  title={Are Benchmark Tests Strong Enough? Mutation-Guided Diagnosis and Augmentation of Regression Suites},
  author={Li, Chenglin and Xu, Yisen and Wang, Zehao and Tan, Shin Hwei and others},
  journal={arXiv preprint arXiv:2604.01518},
  year={2026}
}

@article{DBLP:journals/corr/abs-2511-16004,
  author       = {KeFan Li and
                  Mengfei Wang and
                  Hengzhi Zhang and
                  Zhichao Li and
                  Yuan Yuan and
                  Mu Li and
                  Xiang Gao and
                  Hailong Sun and
                  Chunming Hu and
                  Weifeng Lv},
  title        = {InfCode: Adversarial Iterative Refinement of Tests and Patches for
                  Reliable Software Issue Resolution},
  journal      = {CoRR},
  volume       = {abs/2511.16004},
  year         = {2025},
  url          = {https://doi.org/10.48550/arXiv.2511.16004},
  doi          = {10.48550/ARXIV.2511.16004},
  eprinttype   = {arXiv},
  eprint       = {2511.16004},
  bibsource    = {dblp computer science bibliography, https://dblp.org}
}

@inproceedings{DBLP:conf/iclr/0001LSXTZPSLSTL25,
  author       = {Xingyao Wang and
                  Boxuan Li and
                  Yufan Song and
                  Frank F. Xu and
                  Xiangru Tang and
                  Mingchen Zhuge and
                  Jiayi Pan and
                  Yueqi Song and
                  Bowen Li and
                  Jaskirat Singh and
                  Hoang H. Tran and
                  Fuqiang Li and
                  Ren Ma and
                  Mingzhang Zheng and
                  Bill Qian and
                  Yanjun Shao and
                  Niklas Muennighoff and
                  Yizhe Zhang and
                  Binyuan Hui and
                  Junyang Lin and
                  et al.},
  title        = {OpenHands: An Open Platform for {AI} Software Developers as Generalist
                  Agents},
  booktitle    = {The Thirteenth International Conference on Learning Representations,
                  {ICLR} 2025, Singapore, April 24-28, 2025},
  publisher    = {OpenReview.net},
  year         = {2025},
  url          = {https://openreview.net/forum?id=OJd3ayDDoF},
  bibsource    = {dblp computer science bibliography, https://dblp.org}
}

@inproceedings{DBLP:conf/iclr/JimenezYWYPPN24,
  author       = {Carlos E. Jimenez and
                  John Yang and
                  Alexander Wettig and
                  Shunyu Yao and
                  Kexin Pei and
                  Ofir Press and
                  Karthik R. Narasimhan},
  title        = {SWE-bench: Can Language Models Resolve Real-world Github Issues?},
  booktitle    = {The Twelfth International Conference on Learning Representations,
                  {ICLR} 2024, Vienna, Austria, May 7-11, 2024},
  year         = {2024},
  url          = {https://openreview.net/forum?id=VTF8yNQM66},
  bibsource    = {dblp computer science bibliography, https://dblp.org}
}

@inproceedings{mundler2024swt,
  author       = {Niels M{\"{u}}ndler and
                  Mark Niklas M{\"{u}}ller and
                  Jingxuan He and
                  Martin T. Vechev},
  editor       = {Amir Globersons and
                  Lester Mackey and
                  Danielle Belgrave and
                  Angela Fan and
                  Ulrich Paquet and
                  Jakub M. Tomczak and
                  Cheng Zhang},
  title        = {SWT-Bench: Testing and Validating Real-World Bug-Fixes with Code Agents},
  booktitle    = {Advances in Neural Information Processing Systems 38: Annual Conference
                  on Neural Information Processing Systems 2024, NeurIPS 2024, Vancouver,
                  BC, Canada, December 10 - 15, 2024},
  year         = {2024},
  url          = {http://papers.nips.cc/paper\_files/paper/2024/hash/94f093b41fc2666376fb1f667fe282f3-Abstract-Conference.html},
  bibsource    = {dblp computer science bibliography, https://dblp.org}
}

@inproceedings{DBLP:conf/icse/KangYY23,
  author       = {Sungmin Kang and
                  Juyeon Yoon and
                  Shin Yoo},
  title        = {Large Language Models are Few-shot Testers: Exploring LLM-based General
                  Bug Reproduction},
  booktitle    = {45th {IEEE/ACM} International Conference on Software Engineering,
                  {ICSE} 2023, Melbourne, Australia, May 14-20, 2023},
  pages        = {2312--2323},
  year         = {2023},
  url          = {https://doi.org/10.1109/ICSE48619.2023.00194},
  doi          = {10.1109/ICSE48619.2023.00194},
  bibsource    = {dblp computer science bibliography, https://dblp.org}
}

@article{khatib2025assertflip,
  author       = {Lara Khatib and
                  Noble Saji Mathews and
                  Meiyappan Nagappan},
  title        = {AssertFlip: Reproducing Bugs via Inversion of LLM-Generated Passing
                  Tests},
  journal      = {CoRR},
  volume       = {abs/2507.17542},
  year         = {2025},
  url          = {https://doi.org/10.48550/arXiv.2507.17542},
  doi          = {10.48550/ARXIV.2507.17542},
  eprinttype   = {arXiv},
  eprint       = {2507.17542},
  bibsource    = {dblp computer science bibliography, https://dblp.org}
}

@article{ahmed2025heterogeneous,
  title={Heterogeneous Prompting and Execution Feedback for SWE Issue Test Generation and Selection},
  author={Ahmed, Toufique and Ganhotra, Jatin and Shinnar, Avraham and Hirzel, Martin},
  journal={arXiv preprint arXiv:2508.06365},
  year={2025}
}

@misc{logicstar2025swtbench,
  title        = {{LogicStar} on Test Generation Benchmark {SWT-Bench} {Verified}: Best Test Generation at 84\%},
  author       = {{LogicStar AI}},
  year         = {2025},
  month        = sep,
  howpublished = {LogicStar AI Blog},
  note         = {\url{https://logicstar.ai/blog/logicstar-on-test-generation-benchmark-swt}, accessed 2026-04-27}
}

@article{soni2026swe,
  author       = {Aditya Bharat Soni and
                  Rajat Ghosh and
                  Vaishnavi Bhargava and
                  Valerie Chen and
                  Debojyoti Dutta},
  title        = {SWE-Tester: Training Open-Source LLMs for Issue Reproduction in Real-World
                  Repositories},
  journal      = {CoRR},
  volume       = {abs/2601.13713},
  year         = {2026},
  url          = {https://doi.org/10.48550/arXiv.2601.13713},
  doi          = {10.48550/ARXIV.2601.13713},
  eprinttype   = {arXiv},
  eprint       = {2601.13713},
  bibsource    = {dblp computer science bibliography, https://dblp.org}
}

@article{li2025cocoevo,
  title={CoCoEvo: Co-Evolution of Programs and Test Cases to Enhance Code Generation},
  author={Li, Kefan and Yuan, Yuan and Yu, Hongyue and Guo, Tingyu and Cao, Shijie},
  journal={IEEE Transactions on Evolutionary Computation},
  year={2025},
  publisher={IEEE}
}

@inproceedings{ravi2025llmloop,
  author       = {Ravin Ravi and
                  Dylan Bradshaw and
                  Stefano Ruberto and
                  Gunel Jahangirova and
                  Valerio Terragni},
  title        = {{LLMLOOP:} Improving LLM-Generated Code and Tests Through Automated
                  Iterative Feedback Loops},
  booktitle    = {{IEEE} International Conference on Software Maintenance and Evolution,
                  {ICSME} 2025, Auckland, New Zealand, September 7-12, 2025},
  pages        = {930--934},
  publisher    = {{IEEE}},
  year         = {2025},
  url          = {https://doi.org/10.1109/ICSME64153.2025.00109},
  doi          = {10.1109/ICSME64153.2025.00109},
  bibsource    = {dblp computer science bibliography, https://dblp.org}
}

@article{DBLP:journals/corr/abs-2603-00520,
  author       = {Boxi Yu and
                  Yang Cao and
                  Yuzhong Zhang and
                  Liting Lin and
                  Junjielong Xu and
                  Zhiqing Zhong and
                  Qinghua Xu and
                  Guancheng Wang and
                  Jialun Cao and
                  Shing{-}Chi Cheung and
                  Pinjia He and
                  Lionel C. Briand},
  title        = {{SWE-ABS:} Adversarial Benchmark Strengthening Exposes Inflated Success
                  Rates on Test-based Benchmark},
  journal      = {CoRR},
  volume       = {abs/2603.00520},
  year         = {2026},
  url          = {https://doi.org/10.48550/arXiv.2603.00520},
  doi          = {10.48550/ARXIV.2603.00520},
  eprinttype   = {arXiv},
  eprint       = {2603.00520},
  bibsource    = {dblp computer science bibliography, https://dblp.org}
}

@article{DBLP:journals/corr/abs-2410-06992,
  author       = {Reem Aleithan and
                  Haoran Xue and
                  Mohammad Mahdi Mohajer and
                  Elijah Nnorom and
                  Gias Uddin and
                  Song Wang},
  title        = {SWE-Bench+: Enhanced Coding Benchmark for LLMs},
  journal      = {CoRR},
  volume       = {abs/2410.06992},
  year         = {2024},
  url          = {https://doi.org/10.48550/arXiv.2410.06992},
  doi          = {10.48550/ARXIV.2410.06992},
  eprinttype   = {arXiv},
  eprint       = {2410.06992},
  bibsource    = {dblp computer science bibliography, https://dblp.org}
}

@inproceedings{DBLP:conf/kbse/Soto19,
  author       = {Mauricio Soto},
  title        = {Improving Patch Quality by Enhancing Key Components of Automatic Program
                  Repair},
  booktitle    = {34th {IEEE/ACM} International Conference on Automated Software Engineering,
                  {ASE} 2019, San Diego, CA, USA, November 11-15, 2019},
  pages        = {1230--1233},
  publisher    = {{IEEE}},
  year         = {2019},
  url          = {https://doi.org/10.1109/ASE.2019.00147},
  doi          = {10.1109/ASE.2019.00147},
  bibsource    = {dblp computer science bibliography, https://dblp.org}
}

@misc{copilot,
  author       = {{GitHub}},
  title        = {{GitHub Copilot}},
  year         = {2025},
  howpublished = {\url{https://github.com/features/copilot}},
  note         = {Accessed 2025},
}

@misc{claudecode,
  author       = {{Anthropic}},
  title        = {{Claude Code}},
  year         = {2025},
  howpublished = {\url{https://www.anthropic.com/claude-code}},
  note         = {Accessed 2025},
}

@article{DBLP:journals/computer/DeMilloLS78,
  author       = {Richard A. DeMillo and
                  Richard J. Lipton and
                  Frederick G. Sayward},
  title        = {Hints on Test Data Selection: Help for the Practicing Programmer},
  journal      = {Computer},
  volume       = {11},
  number       = {4},
  pages        = {34--41},
  year         = {1978},
  url          = {https://doi.org/10.1109/C-M.1978.218136},
  doi          = {10.1109/C-M.1978.218136},
  bibsource    = {dblp computer science bibliography, https://dblp.org}
}

@article{DBLP:journals/tse/JiaH11,
  author       = {Yue Jia and
                  Mark Harman},
  title        = {An Analysis and Survey of the Development of Mutation Testing},
  journal      = {{IEEE} Trans. Software Eng.},
  volume       = {37},
  number       = {5},
  pages        = {649--678},
  year         = {2011},
  url          = {https://doi.org/10.1109/TSE.2010.62},
  doi          = {10.1109/TSE.2010.62},
  bibsource    = {dblp computer science bibliography, https://dblp.org}
}

@article{DBLP:journals/corr/abs-2602-10471,
  author       = {Steven Liu and
                  Jane Luo and
                  Xin Zhang and
                  Aofan Liu and
                  Hao Liu and
                  Jie Wu and
                  Ziyang Huang and
                  Yangyu Huang and
                  Yu Kang and
                  Scarlett Li},
  title        = {TestExplora: Benchmarking LLMs for Proactive Bug Discovery via Repository-Level
                  Test Generation},
  journal      = {CoRR},
  volume       = {abs/2602.10471},
  year         = {2026},
  url          = {https://doi.org/10.48550/arXiv.2602.10471},
  doi          = {10.48550/ARXIV.2602.10471},
  eprinttype   = {arXiv},
  eprint       = {2602.10471},
  bibsource    = {dblp computer science bibliography, https://dblp.org}
}

\newpage
\appendix


\section{Empirical Study Details}
\label{app:empirical_setup}

This appendix gives implementation details for the empirical
study of \Cref{sec:empirical}: the mutation patch pool used as
the evaluation instrument (\Cref{app:mpe_pool}), the standalone
BRT-injection study and cutoff-sensitivity check
(\Cref{app:rq1_setup}), and the cogeneration study
(\Cref{app:rq2_setup}).

\subsection{Mutation Patch Evaluation: operators and prompts}
\label{app:mpe_pool}

This appendix details the mutation patch pool used as the
evaluation instrument in \Cref{sec:rq1}.
Given a generated test $t_{\text{gen}}$, we
construct a pool $\mathcal{C}=\{c_1,\dots,c_K\}$ of
plausible-but-incorrect repairs, and use the joint pass/fail
outcomes of $t_{\text{gen}}$ and $t_{\text{golden}}$ on
$\mathcal{C}$ to populate the $2{\times}2$ confusion matrix
that classifies $t_{\text{gen}}$ via \mpe{}'s three-way partition.

\begin{figure}[!t]
  \centering
  \includegraphics[width=0.95\linewidth]{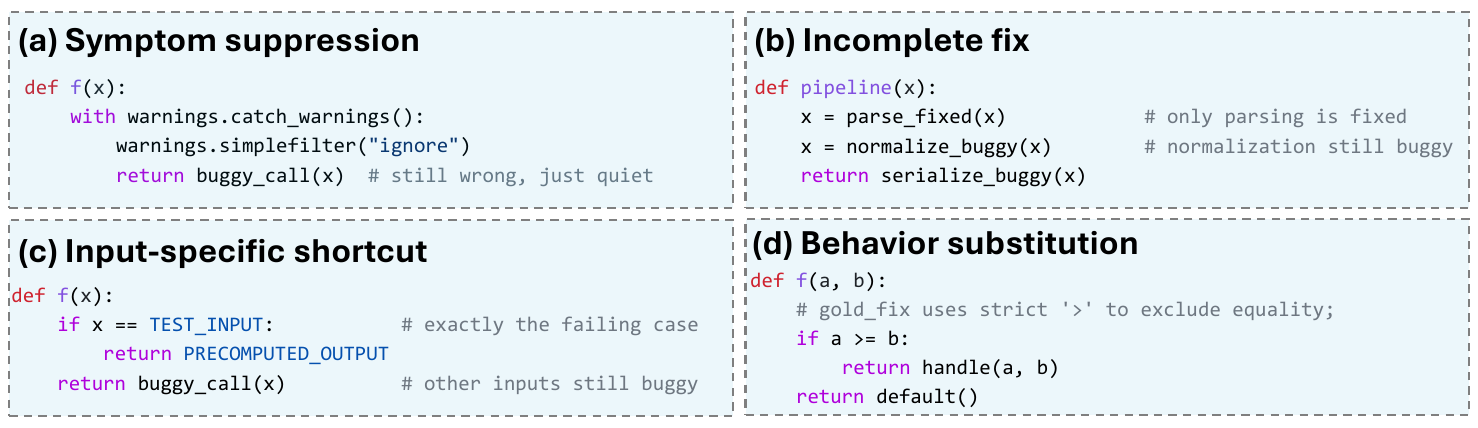}
  \caption{Concrete examples of the four targeted \mpe{}
  operators.
  \textbf{(a) Symptom suppression}: wrapping the buggy call in a
  warning-suppression context lets the bug-introduced warning
  disappear from the test's view.
  \textbf{(b) Incomplete fix}: a multi-stage pipeline is
  repaired only at the parsing stage, leaving normalization and
  serialization buggy.
  \textbf{(c) Input-specific shortcut}: an early branch returns
  a precomputed answer for the test's specific input, bypassing
  the buggy code path.
  \textbf{(d) Behavior substitution}: a comparison operator is
  replaced (\texttt{>=} for \texttt{>}), satisfying tests that
  only check the boundary case qualitatively.}
  \label{fig:operators}
\end{figure}

\smallskip\noindent\textbf{Operators.}
Each mutation patch is generated under one of five
\emph{semantic operators}, targeting common ways an incorrect
fix can pass a reproduction test.
\Cref{fig:operators} illustrates one concrete realization for
each of the four targeted operators.

\begin{itemize}[leftmargin=*]
  \item \textbf{Vanilla.}
    A free-form repair attempt with no operator-specific guidance, used
    as a reference distribution of unconstrained mutation patches.
  \item \textbf{Symptom suppression.}
    Targets tests whose detection signal is an exception or
    warning. The fix does not address the underlying bug;
    instead, it suppresses, swallows, or replaces the signal
    the test inspects.
    Typical realizations: wrap the buggy region in a \texttt{try/except}
    that swallows the exception; suppress warnings via
    \texttt{warnings.catch\_warnings()}; silence error output; catch
    the bug's exception and re-raise a different (expected) type.
  \item \textbf{Incomplete fix.}
    Targets tests that exercise only a strict subset of the
    bug-affected paths (one type, one file, one pipeline stage,
    one caller). Applying the golden fix only to that subset is
    enough to pass the test, while the rest of the buggy paths
    remain unfixed.
    Typical realizations: when $c_{\text{golden}}$ modifies multiple
    locations, edit only one; when $c_{\text{golden}}$ handles multiple
    conditions or types, add an \texttt{if} guard that restricts the
    fix to one of them; when $c_{\text{golden}}$ touches multiple stages
    of a pipeline, only repair one stage.
  \item \textbf{Input-specific shortcut.}
    Targets tests that pin a specific input shape, type, or
    value. A pre-computed shortcut or guard for that input lets
    the test pass while the buggy region of the code remains
    untouched.
    Typical realizations: an early return for the test's specific
    input shape, type, or value; a guard clause that routes the test
    inputs to a clean code path; a cache or memoization layer that
    returns a precomputed result; a special-case branch inserted before
    the buggy region.
  \item \textbf{Behavior substitution.}
    Targets tests that check only partial output properties
    (e.g., output differs from the buggy string, type, shape, or
    finiteness). Replacing the golden fix's core expression or
    call with a syntactically valid alternative satisfies the
    partial check while the resulting behavior is still
    incorrect.
    Typical realizations: a different (e.g., narrower or
    platform-dependent) type cast; a different but similar API call;
    a different comparison operator (e.g., \texttt{>=} for \texttt{>});
    a different ordering of operations.
\end{itemize}

For each instance we sample $56$ patches per operator, yielding
$|\mathcal{C}|=280$ mutation patches per instance.
Patches that fail to apply or fail to produce a runnable repository
are discarded and resampled from the same operator until the budget is
exhausted.

\smallskip\noindent\textbf{Patch generator.}
Mutation patches are generated using the Agentless
pipeline~\citep{DBLP:journals/pacmse/XiaDDZ25} backed by GPT-5-mini. We
prepend the operator-specific instruction to Agentless's
fix-generation prompt and inject three additional context
items: the golden fix $c_{\text{golden}}$, the golden test
$t_{\text{golden}}$, and the generated test $t_{\text{gen}}$.
The two tests are shown as an anonymized pair (Test 1/Test 2,
randomized order) so the mutator is not told which one is generated.
The operator-specific instruction is the only signal that
biases the pipeline toward a particular failure mode.


\subsection{Standalone BRT Injection Study}
\label{app:rq1_setup}

\smallskip\noindent\textbf{Setup.}
We use SWT-Bench~\citep{mundler2024swt} as the instance pool.
For each of the four BRT generators considered---LogicStar~\citep{logicstar2025swtbench},
e-Otter++~\citep{ahmed2025heterogeneous},
AssertFlip~\citep{khatib2025assertflip}, and OpenHands
(\brtonly{} mode)~\citep{DBLP:conf/iclr/0001LSXTZPSLSTL25}---we
download each generator's published BRT predictions from the
prediction logs linked in their GitHub repositories. We then
label the resulting (instance, BRT) pair as \ftp{} or non-\ftp{}
according to whether the test fails on the buggy code and
passes on the golden fix $c_{\text{golden}}$. Pairs on which the
test harness errors out (timeouts, parser failures,
non-importable test modules) are excluded; the remaining pairs
form the two evaluation pools.

\smallskip\noindent\textbf{Run protocol.}
The fix agent is OpenHands~\citep{DBLP:conf/iclr/0001LSXTZPSLSTL25}
backed by GPT-5-mini-0807, with one trial per instance and a
single shared no-test baseline (the same fix agent run on each
instance with no BRT injected).

\smallskip\noindent\textbf{Per-agent pool sizes.}
\Cref{tab:rq1_pool} reports the per-agent pool sizes.
The \ftp{} pool aggregates to $1{,}104$ pairs and the
non-\ftp{} pool to $474$ pairs.
The four agents differ widely in \ftp{} rate: LogicStar covers
$87.1\%$ of instances, while AssertFlip covers $45.7\%$; the
non-\ftp{} pool is correspondingly skewed toward the
lower-coverage agents.
This skew matters when interpreting the per-agent breakdown in
\Cref{tab:rq1_per_agent}, because \ftp{} rate alone does not
predict \tdd utility---LogicStar combines the highest \ftp{}
rate with the largest \tdd gain, whereas AssertFlip combines a
much lower \ftp{} rate with zero gain.

\begin{table}[h]
  \centering
  \small
  \caption{Per-agent pool sizes for the \tdd experiment on
    SWT-Bench.
    The \ftp{} pool is the set of instance-agent pairs on which the
    BRT fails on the buggy code and passes on the golden fix
    $c_{\text{golden}}$; the non-\ftp{} pool is the complementary set,
    after excluding pairs on which the test harness errors out.}
  \label{tab:rq1_pool}
  \begin{tabular}{lrrr}
    \toprule
    Agent & \ftp{} pairs & non-\ftp{} pairs & \ftp{} rate \\
    \midrule
    LogicStar           & 377     & 53  & 87.1\% \\
    e-Otter++           & 272     & 154 & 62.8\% \\
    AssertFlip          & 198     & 127 & 45.7\% \\
    OpenHands-GPT5mini  & 257     & 140 & 59.4\% \\
    \midrule
    Total               & 1{,}104 & 474 & 63.7\% \\
    \bottomrule
  \end{tabular}
\end{table}

\begin{table}[!ht]
  \centering
  \small
  \caption{Per-method breakdown of the RQ1 \tdd experiment under the
    three-way taxonomy of generated BRTs.
    \emph{Overall \tdd{} Resolved} is the fraction of the
    method's instances on which the OpenHands+GPT-5-mini fix agent,
    guided by the injected BRT, produces a fix passing the gold
    tests. Rigorous and Lax partition the method's
    \ftp{} pool by the median of $\beta/\mathrm{total}$ ($=0.202$);
    Misaligned collects tests that reject the golden fix.
    Methods are listed in decreasing order of \ftp{} rate
    (per \Cref{tab:rq1_pool}).}
  \label{tab:rq1_per_agent}
  \begin{tabular}{lrrrrr}
    \toprule
    Test class & \# BRT & \fixonly & \tdd & $\Delta$ & $\Delta_R$ \\
    \midrule
    \multicolumn{6}{l}{\emph{LogicStar~\citep{logicstar2025swtbench} \quad Overall \tdd{} Resolved $=65.8\%$}} \\
    \midrule
    Rigorous & 196 & 61.7\% & 75.5\% & $+$13.8 & $+$36\% \\
    Lax              & 181 & 68.9\% & 69.1\% & $+$0.2 & $+$1\%  \\
    Misaligned       &  53 & 28.3\% & 18.9\% & $-$9.4 & $-$13\% \\
    \midrule
    \multicolumn{6}{l}{\emph{e-Otter++~\citep{ahmed2025heterogeneous} \quad Overall \tdd{} Resolved $=60.3\%$}} \\
    \midrule
    Rigorous & 156 & 69.5\% & 75.9\% & $+$6.4  & $+$21\% \\
    Lax              & 116 & 69.1\% & 65.7\% & $-$3.4  & $-$11\% \\
    Misaligned       & 154 & 45.5\% & 41.6\% & $-$3.9  & $-$7\%  \\
    \midrule
    \multicolumn{6}{l}{\emph{OpenHands~\citep{DBLP:conf/iclr/0001LSXTZPSLSTL25} \quad Overall \tdd{} Resolved $=61.0\%$}} \\
    \midrule
    Rigorous & 123 & 71.2\% & 76.1\% & $+$4.9  & $+$17\% \\
    Lax              & 134 & 72.5\% & 74.7\% & $+$2.2  & $+$8\%  \\
    Misaligned       & 140 & 40.0\% & 40.7\% & $+$0.7  & $+$1\%  \\
    \midrule
    \multicolumn{6}{l}{\emph{AssertFlip~\citep{khatib2025assertflip} \quad Overall \tdd{} Resolved $=58.9\%$}} \\
    \midrule
    Rigorous &  77 & 71.1\% & 76.3\% & $+$5.2  & $+$18\% \\
    Lax              & 121 & 70.0\% & 66.7\% & $-$3.3  & $-$11\% \\
    Misaligned       & 127 & 55.1\% & 42.5\% & $-$12.6 & $-$28\% \\
    \bottomrule
  \end{tabular}
\end{table}

\smallskip\noindent\textbf{Cutoff sensitivity.}
\label{app:cutoff_sensitivity}

\Cref{tab:rq1_tdd} splits the \ftp{} pool by the median
Laxity rate, reported as $\tau=0.202$. To check that the
finding is not an artifact of this cutoff, we sweep $\tau$ while
holding the same $1{,}104$ \ftp{} records fixed.
\Cref{tab:app_cutoff_sweep} reports representative cutoffs. The
Rigorous--Lax gap remains positive even when the partition is far
from balanced.

\begin{table}[H]
  \centering
  \small
  \caption{Cutoff sensitivity of the Rigorous/Lax split. For each
  Laxity-rate cutoff $\tau$, $\Delta_{\text{Rigorous}}$ is the Resolved gain
  from injecting tests with Laxity rate below $\tau$, and
  $\Delta_{\text{Lax}}$ is the corresponding gain for tests at or
  above $\tau$, both measured relative to running the same
  instance without injecting a BRT.}
  \label{tab:app_cutoff_sweep}
  \begin{tabular}{lrrrrr}
    \toprule
    Cutoff $\tau$ & $N_{\text{Rigorous}}$ & $N_{\text{Lax}}$ &
    $\Delta_{\text{Rigorous}}$ & $\Delta_{\text{Lax}}$ & Gap \\
    \midrule
    0.050 & 315 & 789 & +8.6 & +2.5 & +6.0 \\
    0.100 & 403 & 701 & +8.2 & +2.0 & +6.2 \\
    0.150 & 491 & 613 & +8.6 & +0.8 & +7.7 \\
    \textbf{0.202} & \textbf{552} & \textbf{552} &
    \textbf{+8.5} & \textbf{+0.0} & \textbf{+8.5} \\
    0.250 & 610 & 494 & +7.2 & +0.6 & +6.6 \\
    0.300 & 678 & 426 & +7.1 & -0.2 & +7.3 \\
    0.400 & 770 & 334 & +6.0 & +0.3 & +5.7 \\
    0.500 & 844 & 260 & +5.6 & +0.0 & +5.6 \\
    \bottomrule
  \end{tabular}
\end{table}

\subsection{Cogeneration Study}
\label{app:rq2_setup}

\smallskip\noindent\textbf{Setup.}
Our \cogen{} configuration follows the joint fix-and-test design
of~\citet{DBLP:journals/corr/abs-2601-19066}, which has no
official SWT-Bench implementation; we implement it by adapting
the OpenHands~\citep{DBLP:conf/iclr/0001LSXTZPSLSTL25} prompt
template to also emit a reproduction test in the same
trajectory. The configuration uses GPT-5-mini-0807 as the
underlying language model with one trial per instance on
SWT-Bench.

\smallskip\noindent\textbf{Comparison configurations.}
In \cogen{} mode, the agent emits both a fix and a test in a
single trajectory. In \brtonly{} mode, the same agent is asked
only for a reproduction test (fix emission disabled). In
\fixonly{} mode, the agent is asked only for a fix (test
emission disabled). The three configurations share the same
backbone, prompt scaffold, and inference settings; they differ
only in the artifacts requested from the agent within the
trajectory.

\smallskip\noindent\textbf{Joint-failure baseline.}
Let $p_{\text{fix}}$ be the marginal rate of unresolved fixes
($166/433$) and $p_{\text{test}}$ the marginal rate of non-\ftp{}
tests ($120/433$). Under independent fix and test errors, the
joint-failure rate is $p_{\text{fix}}\cdot p_{\text{test}}\!=\!10.6\%$;
the observed joint-failure rate is $86/433\!=\!19.9\%$,
giving a coupling factor of $1.87$.
\section{Experimental Setup Details}
\label{app:setup}
\noindent\textbf{Benchmark.}
We use the 433-instance intersection of SWE-bench Verified and SWT-bench Verified, which enables joint evaluation of patch correctness and test effectiveness. To assess cross-benchmark generalization, we additionally evaluate on SWE-bench Lite. Dataset construction, filtering criteria, and statistics are detailed in \Cref{app:dataset}.

\noindent\textbf{Metrics.}
\textbf{Resolved} measures the fraction of instances where the generated patch passes the SWE-bench evaluation harness. \textbf{\ftp{}} measures the fraction of instances where the generated test fails on the buggy program and passes on the corresponding reference fix, following the SWT-bench protocol. Together, these metrics capture complementary aspects of correctness (fix validity vs.\ test effectiveness).

\noindent\textbf{Baselines.}
We compare against two families: (i) \emph{fix-only} agents that generate patches without explicit reproduction tests, and (ii) \emph{cogeneration} methods that jointly produce patches and tests in a single trajectory. The full list of baselines and their configurations are provided in \Cref{tab:main_fix}, with detailed descriptions in \Cref{app:baselines}.

\noindent\textbf{Implementation.}
We implement \ours{} on top of OpenHands with GPT-5-mini as the backbone LLM. The hardening procedure runs for $K=5$ rounds. In each round, we construct a mutant pool $\mathcal{C}_k$ using the four targeted mutation operators described in \Cref{sec:mpe}, with $N=3$ samples per operator (up to $|\mathcal{C}_k|=12$). All OpenHands-based methods share the same backbone to ensure fair comparison. For non-OpenHands baselines, we report results from their official implementations. Additional implementation details, including prompts and per-round procedures, are provided in \Cref{app:method_impl}.

\subsection{Dataset Details}
\label{app:dataset}


\paragraph{SWE-bench.}
SWE-bench~\citep{DBLP:conf/iclr/JimenezYWYPPN24} collects real-world GitHub issues paired with developer-written patches (\emph{golden fixes}) from 12 popular open-source Python repositories (Django, scikit-learn, sympy, matplotlib, etc.). SWE-bench Verified is a 500-instance human-validated subset whose golden fixes are confirmed correct and self-contained. SWE-bench Lite is a separate 300-instance subset curated for faster evaluation.

\paragraph{SWT-bench.}
SWT-bench~\citep{mundler2024swt} re-purposes the same issue--patch instances but targets \emph{test generation}: each instance is evaluated against a golden fail-to-pass test extracted from the original pull request. Because not every SWE-bench instance has a test patch suitable for fail-to-pass evaluation, SWT-bench filters out a small number of incompatible cases from each SWE-bench subset, 67 from Verified and 24 from Lite, yielding 433 and 276 usable instances respectively.
\paragraph{Evaluation sets.}
Our main experiments (\Cref{sec:rq1_main,sec:rq3_ablation}) use the 433-instance SWE-bench Verified $\cap$ SWT-bench Verified subset, where every instance possesses both a golden fix and a golden test, enabling joint evaluation of fix correctness (Resolved) and test quality (\ftp{}). The cross-benchmark experiment (\Cref{sec:rq2_general}) uses the 276-instance SWE-bench Lite $\cap$ SWT-bench Lite subset.

\subsection{Baseline Implementation Details}
\label{app:baselines}

We compare \ours{} against two families of methods. \textbf{(i)~Fix-only}: agents that generate only a fix, including SWE-Agent and mini SWE-Agent~\citep{DBLP:conf/nips/YangJWLYNP24}, Agentless~\citep{DBLP:journals/pacmse/XiaDDZ25}, and OpenHands~\citep{DBLP:conf/iclr/0001LSXTZPSLSTL25}.
\textbf{(ii)~Cogeneration}: methods that jointly produce the fix and the BRT, including InfCode~\citep{DBLP:journals/corr/abs-2511-16004}, Agent-CoEvo~\citep{li2026beyond}, and OpenHands\,+\,\cogen{} (same backbone as \ours{}). Below we describe each method in detail.

\paragraph{Fix-only.}
Fix-only methods produce a patch without generating an
explicit bug-reproduction test. The fix is evaluated directly
against the golden test suite.
\begin{itemize}
  \item \textbf{SWE-Agent / mini SWE-Agent~\cite{DBLP:conf/nips/YangJWLYNP24}.}
    SWE-Agent equips an LLM with a custom Agent--Computer Interface~(ACI) that exposes repository navigation, file viewing, and editing as LM-friendly commands. The ACI abstractions (e.g., a scrollable file viewer, a targeted edit command with lint feedback) substantially improve the agent's ability to resolve software-engineering issues end-to-end. mini SWE-Agent is a simplified variant ($\sim$100 lines) of SWE-Agent from the same team, designed for minimal complexity while retaining competitive performance. 
  \item \textbf{Agentless~\cite{DBLP:journals/pacmse/XiaDDZ25}.} 
  Agentless follows a non-agentic, three-phase pipeline: (1)~hierarchical fault localization that narrows the search from file to class/function to fine-grained edit location, (2)~LLM-based patch generation via sampling, and (3)~patch selection using regression tests and LLM-generated reproduction tests. Unlike agent-based methods, the LLM does not decide future actions or wield interactive tools; it is queried in a fixed, structured sequence.
  \item \textbf{OpenHands~\cite{DBLP:conf/iclr/0001LSXTZPSLSTL25}.}
    OpenHands is a composable software-agent platform that provides sandboxed runtime environments, an event-stream architecture, and modular agent components for planning, memory, and skill execution. In fix-only mode the agent interactively browses the repository, localizes the bug, and applies a patch without generating a reproduction test.
\end{itemize}

\paragraph{Cogeneration.}
Cogeneration baselines produce the fix and the reproduction
test within a single trajectory, allowing mutual feedback
between the two artifacts.
\begin{itemize}
    \item \textbf{InfCode~\cite{DBLP:journals/corr/abs-2511-16004}.}
    InfCode is a multi-agent adversarial framework comprising a Test Patch Generator, a Code Patch Generator, and a Selector agent. The test agent generates tests that expose failures, the code agent refines patches to pass them, and the Selector picks the best fix. This adversarial co-refinement iterates inside a containerized repository environment. Results are obtained from the official open-source implementation.
    \item \textbf{Agent-CoEvo~\cite{li2026beyond}.}
    Agent-CoEvo frames issue resolution as coevolution: candidate code patches and test patches are jointly explored by multiple agents, with each side mutually evaluating and semantically recombining the other's candidates across iterations. Tests are treated as dynamic behavioral constraints that evolve alongside the code rather than fixed oracles.  Since no official implementation is available, we reproduce it on top of the InfCode codebase from the same authors.
    \item \textbf{OpenHands\,+\,\cogen{}.}
    Dynamic Cogeneration studies workflow orderings for jointly generating a fix and a BRT inside the OpenHands agent. The agent alternates between fix and test steps within one trajectory, using the co-generated test as an in-trajectory F$\to$P gate to guide subsequent iterations.
\end{itemize}

\section{Method implementation}
\label{app:method_section}

\subsection{\ours{} implementation notes}
\label{app:method_impl}

\Cref{alg:diagcoevo} states the full procedure. This appendix
records implementation choices that support the algorithm but do
not affect its algorithmic content. \ours{} ships as a
single-skill instantiation of \Cref{app:skill_critic}:
\texttt{critic.yaml} declares one \texttt{hard\_gate} signal
whose detector implements \Cref{alg:diagcoevo} end-to-end.

\begin{algorithm}[H]
\small
\caption{\ours: iterative-hardening cogeneration.}
\label{alg:diagcoevo}
\begin{algorithmic}[1]
\Require issue $I$; max rounds $K$; mutation operators $\mathcal{O}$; samples per operator $N$; regression tolerance $\tau$ (default $0.2$).
\Ensure reproduction test $t^\star$, fix $c^\star$.
\State \(\triangleright\) \emph{Phase~1 (Round 0): lax-init}
\State $t_0 \gets \textsc{Agent}_{\text{test-only}}(I)$
\State $c_0 \gets \emptyset$;\ $t_{\text{ref}} \gets t_0$;\ $\text{feedback} \gets \emptyset$
\State \(\triangleright\) \emph{Phase~2 (Rounds 1..K): hardening}
\For{$k = 1, \ldots, K$}
  \State $(t_k, c_k) \gets \textsc{Agent}_{\text{joint}}(I,\ t_{\text{ref}},\ c_{k-1},\ \text{feedback})$
  \State $\mathcal{C}_k \gets \bigcup_{o \in \mathcal{O}} \textsc{Mutate}(c_k, o, N)$
  \State $\mathrm{TM}_k \gets \textsc{TemporalMatrix}(t_k,\ t_{\text{ref}},\ \mathcal{C}_k)$ \Comment{cells $\alpha,\beta,\gamma,\delta$}
  \State $\ell_k \gets \beta(\mathrm{TM}_k) \,/\, \mathrm{total}(\mathrm{TM}_k)$ \Comment{Laxity rate}
  \If{$\ell_k > \tau$} \Comment{regression: revert test to $t_0$, keep fix}
    \State $t_k \gets t_0$;\ $t_{\text{ref}} \gets t_0$
  \ElsIf{$t_k \neq t_{\text{ref}}$} \Comment{clean round with non-trivial edit: exit}
    \State \Return $(t_k, c_k)$
  \Else
    \State $t_{\text{ref}} \gets t_k$ \Comment{test unchanged: continue with retry feedback}
  \EndIf
  \State $\text{feedback} \gets \textsc{Render}(\mathrm{TM}_k,\ \mathcal{C}_k)$
\EndFor
\State \Return $(t_K, c_K)$ \Comment{budget exhausted}
\end{algorithmic}
\end{algorithm}

\smallskip\noindent\textbf{Convergence.}
The loop exits on (i) a \emph{clean round}, where the agent's
$t_k$ differs from the reference $t_{\text{ref}}$ and the
Laxity rate $\ell_k = \beta / \mathrm{total}$ stays at or
below tolerance $\tau$, where $\mathrm{total}$ is the number of
admissible mutants in that round, or (ii) the round budget $K$. A round
with $\ell_k > \tau$ instead reverts $t_k$ to the lax-init test
$t_0$ while retaining $c_k$, then continues to round $k+1$ with
feedback that names the regressed mutants. We use $K = 5$ and
$\tau = 0.2$ as defaults in \Cref{sect:exp}.

\smallskip\noindent\textbf{Variant generation budget.}
At each round the loop calls the patch-mutation LLM with $N$
samples per operator
(default $N=3$, total $|\mathcal{C}_k|=12$ when all four targeted
operators are active) in parallel.
Variants that fail to apply, time out, or raise an environment
error are excluded from $|\mathcal{C}_k|$ rather than counted as
survivors.
A round is marked \emph{inconclusive} if zero admissible variants
are produced; the regression check is skipped for that round and
the inconclusive count is tracked separately.

\smallskip\noindent\textbf{State persistence across agent invocations.}
\Cref{alg:diagcoevo} treats the loop as a single program, but in
the deployed system each iteration of the \texttt{for} loop is a
separate agent invocation: the agent calls \texttt{finish}, the
critic evaluates, and on rejection the SDK injects the followup
and lets the agent continue.
A small JSON state object
($t_{\text{prev}}$, round counter, inconclusive
streak, fix history for the previous-round survivor check) is
persisted to disk between invocations and reset whenever the
benchmark instance changes.

\smallskip\noindent\textbf{Sampling budget vs.\ post-hoc evaluation.}
The post-hoc evaluation in \Cref{app:mpe_pool} uses a much larger
pool ($56$ patches per operator, $|\mathcal{C}|=280$) because it
is a one-shot offline measurement; the in-loop pool is
deliberately small ($N=3$) because its cost is paid per agent
round and the loop amortizes the evaluation over $K$ rounds.

\smallskip\noindent\textbf{Cost by hardening round.}
\Cref{tab:app_round_cost} reports cumulative cost in the final
\ours{} run. The number of active instances drops sharply after
R1, and most instances no longer trigger MPE mutant generation in
later rounds. This behavior is intentional. The convergence rule
exits once the current test has changed non-trivially without
introducing Lax regressions, so instances whose test--fix pair is
already sufficiently strong do not pay for unnecessary hardening.
This matches the expected use of \ours{}, which reserves extra rounds for
the harder tail rather than forcing every instance through the full
budget.

\begin{table}[H]
  \centering
  \small
  \caption{Cumulative cost by hardening round. \emph{Active instances}
  counts instances that still trigger MPE mutant generation in that
  round. \emph{Total cost / inst.} is the cumulative end-to-end LLM
  cost averaged over the full 433-instance benchmark. \emph{MPE cost /
  inst.} is the subset of that cost spent on mutation-patch generation,
  also averaged over 433 instances; percentages in parentheses show
  its share of the total cumulative cost in the same row. R5 is the
  final submitted output; no new MPE mutants are generated after R4,
  so MPE cost stays unchanged.}
  \label{tab:app_round_cost}
  \begin{tabular}{lrrr}
    \toprule
    Round & Active instances & Total cost / inst. & MPE cost / inst. \\
    \midrule
    R0 & 433 & \$0.115 & \$0.000 (0.0\%) \\
    R1 & 433 & \$0.492 & \$0.049 (10.0\%) \\
    R2 & 106 & \$0.638 & \$0.062 (9.7\%) \\
    R3 & 80  & \$0.696 & \$0.071 (10.2\%) \\
    R4 & 62  & \$0.744 & \$0.079 (10.6\%) \\
    \textbf{R5 / Final} & 0 & \textbf{\$0.834} & \textbf{\$0.079 (9.5\%)} \\
    \bottomrule
  \end{tabular}
\end{table}

\smallskip\noindent\textbf{Budget-matched self-refinement baseline.}
To check whether the gains come only from extra refinement budget,
we evaluate a pure Ralph-style baseline. Ralph starts from the same
vanilla cogeneration prompt and uses the same $K=5$ round budget,
but receives only generic ``review and improve'' feedback; it does
not use mutation feedback or CoHarden diagnostic checks.

\begin{table}[H]
  \centering
  \small
  \caption{Budget-matched self-refinement baseline. Generic
  self-refinement improves Resolved over vanilla cogeneration, but
  does not match CoHarden and loses test quality.}
  \label{tab:app_ralph_baseline}
  \begin{tabular}{lrr}
    \toprule
    Method & Test F$\to$P (\%) & Resolved (\%) \\
    \midrule
    Vanilla cogeneration ($K=1$) & 72.3 & 61.7 \\
    Pure Ralph self-refine ($K=5$) & 66.7 & 63.8 \\
    \textbf{\ours{} (CoHarden)} & \textbf{78.9} & \textbf{69.4} \\
    \bottomrule
  \end{tabular}
\end{table}

\subsection{\textsc{PluggableCritic} for fine-grained loop control}
\label{app:skill_critic}

\smallskip\noindent\textbf{Background.}
We build on three mechanisms that today's LLM-agent stacks expose
in isolation.
A \emph{Ralph-style loop} is the simplest possible
autonomous-agent driver: a coding tool such as
Claude Code~\citep{claudecode},
GitHub Copilot~\citep{copilot}, or
OpenHands~\citep{DBLP:conf/iclr/0001LSXTZPSLSTL25} is re-invoked
on the same task until it declares completion, in spirit
\begin{verbatim}
  while :; do cat PROMPT.md | claude-code --continue; done
\end{verbatim}
A \emph{skill} is a self-contained, reusable, pluggable unit of
domain knowledge that recent SDKs (Claude Code Skills,
GitHub Copilot's skill packs, the OpenHands skill catalog)
standardize: a skill bundles prompts, reference documents, and
small tools that the agent can read or invoke during a turn.
A \emph{critic} is an agent-internal scoring component that
inspects what the agent has done and decides whether the
submission is acceptable or whether the agent should retry.

\smallskip\noindent\textbf{Where each primitive falls short.}
Composing these three primitives to produce a domain-controlled
loop runs into four gaps.
\textbf{1.} Loops driven by Ralph-style scripts (or any
external orchestrator) only see what the agent prints to
\texttt{stdout} and the process exit code; they cannot inspect
the agent's trajectory (which tools were called, what
intermediate artifacts were produced, whether the agent actually
ran the test) and so cannot make behavior-level decisions.
\textbf{2.} Skills are \emph{passive} from the loop's
perspective: the agent decides when, or whether, to consult
them, and there is no surface to enforce that a skill's
protocol was actually followed before the loop accepts a
submission.
\textbf{3.} There is no community-standard verify-and-feedback
interface; each project hand-rolls its own retry harness,
verification logic gets entangled with task logic, and good
verifiers cannot be packaged and shared the way skills can.
\textbf{4.} Retry feedback in script-driven loops is largely
static (``try again, more carefully''); there is no clean way to
push trajectory-derived dynamic content (which files were
touched, which mutated patches survived, what the per-operator
counts are) back into the next turn.

\smallskip\noindent\textbf{\textsc{PluggableCritic}.}
\textsc{PluggableCritic} combines the pluggability of a skill
with the trajectory access of a critic, and exposes the
combination as a user-defined verify-and-feedback boundary at
the agent's \texttt{finish} action.
On \texttt{finish}, the harness invokes the user-supplied critic,
which inspects the agent's full event stream (tool calls,
observations, messages) plus the working-tree diff and returns a
score in $[0,1]$.
If the score falls below a configured threshold, the harness
injects the critic's followup prompt as the next user turn and
lets the agent retry, up to a configured maximum.
This addresses gap 1 (the critic receives the full event stream
and the working-tree diff), gap 2 (the critic can require
arbitrary post-conditions before the loop accepts), gap 3 (the
critic itself ships as a swappable unit, configured per
experiment), and gap 4 (the followup prompt is composed by the
critic and can carry trajectory-derived evidence).

\smallskip\noindent\textbf{Instantiation.}
A user instantiates a \textsc{PluggableCritic} in one of two
complementary modes. 
The \emph{subclass mode} requires implementing one method on the
abstract base \texttt{CriticBase}:
\texttt{evaluate(events, git\_patch) -> CriticResult}, where
\texttt{CriticResult} carries a score, a message, and a free-form
metadata dictionary;
\texttt{get\_followup\_prompt(result, iteration) -> str} is
optionally overridden to compose dynamic feedback from that
metadata (the default returns a generic ``please try again''
message).
The \emph{skill mode}, used by \ours{}, instead loads the critic
declaratively from a skill directory: a \texttt{critic.yaml}
manifest lists named signals (each typed as \texttt{hard\_gate}
or \texttt{soft} with a weight) together with their detector
file, and each detector is a small Python module exposing
\texttt{detect(events, git\_patch) -> (triggered, metadata)}.
The harness short-circuits on the first triggered hard gate and
otherwise sums the soft-signal weights against the threshold; the
default followup prompt walks the triggered signals and
concatenates each detector's
\texttt{metadata["followup\_text"]}.
The \texttt{git\_patch} argument is the agent's working-tree diff
at \texttt{finish} time, computed by the harness so detectors do
not each have to recompute it.
Two implementation choices support \ours{} specifically:
detectors run in-process via \texttt{importlib} so they share the
SDK's LLM client (variant generation and survivor analysis would
otherwise pay a serialization cost per round), and detectors own
their feedback text (sampled survivor patches and per-operator
counts) so dynamic content flows back to the agent without an
SDK-side templating layer.
The current implementation is coupled to the OpenHands SDK's
event and tool model; the same design admits a small adapter
layer for any other skill-based SDK that exposes a
\texttt{finish}-equivalent boundary.



\smallskip\noindent\textbf{Reusability and release.}
The contract is intentionally minimal, so the same harness hosts
critics that have nothing to do with \ours{}: a clean-patch gate
(rejects non-test files in test-only phases), a
structural-validity gate (rejects unparseable diffs), and the
test-only enforcement gate used by the \brtonly{} baseline of
\Cref{sec:rq2} all ship as separate single-detector skills
without code changes to the harness.

\subsection{Computational Resources and Software Environment}
\label{app:compute}

\smallskip\noindent\textbf{Hardware.}
All experiments were conducted on a single server running
\textbf{Ubuntu 24.04.1 LTS}, equipped with four Intel Xeon
E7-4830\,v3 processors (12 cores per socket, 48 physical cores,
96 logical threads at 2.10\,GHz) and 503\,GB of RAM. No GPU was
used: all workloads are LLM API calls, Docker-based test
execution, and lightweight scripting.

\smallskip\noindent\textbf{Software environment.}
The orchestration codebase is written in \textbf{Python~3.13}
managed via \textbf{uv}. Key dependencies include
OpenHands SDK, SWE-agent, Agentless, and the OpenAI Python
client (\texttt{openai~$\geq$~2.24}). Each SWE-bench instance runs inside an isolated Docker
container with its own Conda environment, following the
SWE-bench harness protocol.

\smallskip\noindent\textbf{LLM API models.}
All LLM calls are made through commercial APIs.
The per-token pricing (per 1\,M tokens) is:
GPT-5-mini (\$0.25\,/\,\$2.00 input\,/\,output),
GPT-5.4 (\$2.50\,/\,\$15.00), and
Claude Opus 4.5 (\$5.00\,/\,\$25.00).
All cost estimates assume \textbf{zero prompt-cache hit rate}
to ensure a fair, reproducible upper bound.





\newpage
\section{Prompts}
\label{appendix: prompts}

In this section, we provide the prompts used in the different configurations of our framework.


\begin{tcolorbox}[
  breakable,
  enhanced,
  colback=lightgray!20,
  colframe=darkgray!80,
  title=Mutation Operator Definitions
]
\label{app:prompt_operators}
\small

Four semantic-level mutation operators used to generate adversarial code patches. Each operator instructs the LLM to produce a ``plausible but incorrect'' fix using a specific strategy. These operators are injected into the sampling prompt via the \texttt{\{\{ operator \}\}} variable.

\smallskip
\textbf{Operator 1: Symptom Suppression}

\smallskip
The bug still occurs and produces incorrect behavior, but you HIDE the observable signal that the test relies on to detect it. The test checks for a specific error signal (exception, warning, assertion error, wrong error message, etc.). Your patch should suppress or silence that signal without fixing the underlying computation.

\smallskip
Common techniques:
\begin{itemize}[leftmargin=*,noitemsep]
    \item Wrap buggy code in try/except to swallow exceptions
    \item Use \texttt{warnings.catch\_warnings()} to suppress warnings
    \item Redirect or silence error output
    \item Catch and re-raise a different (expected) exception type
\end{itemize}

\smallskip
\textbf{Operator 2: Incomplete Fix}

\smallskip
Take the gold fix and REMOVE part of it. Apply only a SUBSET of the changes --- enough for the test to pass, but leaving some of the bug unfixed.

\smallskip
Concrete operations:
\begin{itemize}[leftmargin=*,noitemsep]
    \item If the gold fix modifies multiple locations, apply changes at only ONE location and leave the others unchanged
    \item If the gold fix handles multiple conditions/types, add an \texttt{if} guard that restricts it to only ONE condition/type
    \item If the gold fix changes multiple stages of a pipeline, only fix one stage and leave the rest broken
    \item If the gold fix adds both a computation AND error handling, only add the computation and skip the error handling
\end{itemize}

\smallskip
\textbf{Operator 3: Input-Specific Shortcut}

\smallskip
You add a SHORTCUT that prevents the test input from ever reaching the buggy code. The bug is not fixed --- the buggy code still exists unchanged --- but the test input is intercepted and handled separately before it gets there.

\smallskip
Common techniques:
\begin{itemize}[leftmargin=*,noitemsep]
    \item Add an early return for the specific input shape/type/value
    \item Add a guard clause that routes test inputs to a clean path
    \item Add a cache/memoization that returns a pre-computed result
    \item Insert a special-case branch before the buggy code region
\end{itemize}

\smallskip
\textbf{Operator 4: Behavior Substitution}

\smallskip
Take the gold fix and REPLACE its core expression or function call with a DIFFERENT one that is syntactically valid and plausible. Keep the same code location but use a different approach.

\smallskip
Key constraints:
\begin{itemize}[leftmargin=*,noitemsep]
    \item Your patch must touch the SAME lines as the gold fix --- do not add code elsewhere
    \item Replace the key expression/operation with a different but syntactically valid alternative
\end{itemize}

\smallskip
Concrete operations:
\begin{itemize}[leftmargin=*,noitemsep]
    \item If the gold fix uses one type cast, use a DIFFERENT type cast (e.g., a narrower or platform-dependent type)
    \item If the gold fix uses one API/function, use a DIFFERENT API/function that has similar but not identical behavior
    \item If the gold fix uses one comparison operator, use a DIFFERENT operator (e.g., \texttt{>=} instead of \texttt{>})
    \item If the gold fix reorders operations in one way, reorder them in a DIFFERENT way
\end{itemize}

\end{tcolorbox}

\newpage

\begin{tcolorbox}[
  breakable,
  enhanced,
  colback=lightgray!20,
  colframe=darkgray!80,
  title=Mutation Patch Sampling Prompt
]
\label{app:prompt_sampling}
\small

We are currently solving the following issue within our repository. Here is the issue text:\\
\texttt{-{-}- BEGIN ISSUE -{-}-}\\
\texttt{\{\{ problem\_statement \}\}}\\
\texttt{-{-}- END ISSUE -{-}-}

\smallskip
Below are some code segments, each from a relevant file. One or more of these files may contain bugs.\\
\texttt{-{-}- BEGIN FILE -{-}-}\\
\texttt{\{\{ content \}\}}\\
\texttt{-{-}- END FILE -{-}-}

\smallskip
\texttt{\{\{ operator \}\}}

\smallskip
Generate \texttt{edit\_file} commands to implement your patch.

\smallskip
\texttt{-{-}- Test 1 -{-}-}\\
\texttt{\{\{ test\_1 \}\}}\\
\texttt{-{-}- End Test 1 -{-}-}

\smallskip
\texttt{-{-}- Test 2 -{-}-}\\
\texttt{\{\{ test\_2 \}\}}\\
\texttt{-{-}- End Test 2 -{-}-}

\smallskip
Two tests exist for this bug. Study both carefully and note their differences in assertions, inputs, and coverage. Your patch should produce DIFFERENT outcomes on these two tests --- passing one while failing the other.

\smallskip
\texttt{-{-}- Reference Fix (the correct developer patch) -{-}-}\\
\texttt{\{\{ gold\_fix \}\}}\\
\texttt{-{-}- End Reference Fix -{-}-}

\smallskip
The above is the correct fix. Use it as a starting point to generate a VARIANT patch --- one that partially fixes the bug or fixes it in a subtly different way, so that the two tests above produce different outcomes.

\smallskip
The \texttt{edit\_file} command takes four arguments:

\smallskip
\texttt{edit\_file(filename: str, start: int, end: int, content: str) -> None:}\\
\hspace*{1em}Edit a file. It replaces lines \texttt{start} through \texttt{end} (inclusive) with the given text \texttt{content} in the open file.\\
\hspace*{1em}Args:\\
\hspace*{2em}\texttt{filename}: The full file name to edit.\\
\hspace*{2em}\texttt{start}: The start line number. Must satisfy \texttt{start >= 1}.\\
\hspace*{2em}\texttt{end}: The end line number. Must satisfy \texttt{start <= end <= number of lines in the file}.\\
\hspace*{2em}\texttt{content}: The content to replace the lines with.

\smallskip
Please note that THE \texttt{edit\_file} FUNCTION REQUIRES PROPER INDENTATION. If you would like to add the line `\texttt{~~~~~~~~print(x)}', you must fully write that out, with all those spaces before the code! Wrap the \texttt{edit\_file} command in blocks \texttt{```python...```}.

\end{tcolorbox}

\newpage

\begin{tcolorbox}[
  breakable,
  enhanced,
  colback=lightgray!20,
  colframe=darkgray!80,
  title=\ours{} Lax-Init Prompt (Round 0)
]
\label{app:prompt_laxinit}
\small

\texttt{<uploaded\_files>}\\
\texttt{/workspace/\{\{ workspace\_dir\_name \}\}}\\
\texttt{</uploaded\_files>}

\smallskip
I've uploaded a python code repository in the directory \texttt{\{\{ workspace\_dir\_name \}\}}. Consider the following issue description:

\smallskip
\texttt{<issue\_description>}\\
\texttt{\{\{ instance.problem\_statement \}\}}\\
\texttt{</issue\_description>}

\smallskip
\{\% block task \%\}\\
Your task is to write a test that reproduces this bug, then fix it.\\
\{\% endblock \%\}

\smallskip
Your fix should resolve the issue. Your test should reproduce the issue --- fail before the fix, pass after. Keep the test implementation-agnostic: verify behavior from the project's semantics, not details specific to your fix.

\smallskip
After each step you will receive feedback guiding you to the next step. The workflow is:
\begin{itemize}[leftmargin=*,noitemsep]
    \item \textbf{Round 0}: write \texttt{reproduction.py} only --- no fix yet.
    \item \textbf{Round 1}: find the root cause and the boundary.
    \item \textbf{Round 2+} (Hardening): refine both the fix and the test based on feedback.
\end{itemize}

\smallskip
\textbf{Guidelines for Round 0}:

\smallskip
The issue description mixes two kinds of information: \textbf{OBSERVATION} (pasted code, exact inputs, printed strings, error messages, stack traces, or the wrong value that came out) and \textbf{ATTRIBUTION} (claims about what should happen or which component is at fault). Reuse observed inputs verbatim as the reproduction fixture, but do not encode attribution directly as the oracle unless you independently verify it from project semantics.

\smallskip
Your reproduction distinguishes two world states: bug present (exit non-zero) vs.\ bug absent (exit zero). A script that reports ``no bug detected'' on fixed code is a successful reproduction. Write the bug-present predicate $P$ explicitly and use the fixed shape \texttt{sys.exit(1 if P else 0)}.

\smallskip
\textbf{What to put in $P$}: if the reporter's expected value is verifiable from a stable independent source (mathematical/library specification, existing tests, docstrings, or documented public API), derive it yourself and set $P$ to detect deviation from that value. If it is not verifiable, use an invariant that any correct fix must preserve, such as difference across feature settings, equivalence to an independent implementation, round-trip preservation, or absence of an exception on valid input. For rendered text output, parse into the semantic object when possible rather than substring-matching raw formatting.

\smallskip
In Round 0, make the final user-visible observable your \textbf{main witness}: printed output, rendered response, exception text, returned value, or visible side effect. Use internal values only as supporting sanity checks.

\smallskip
\textbf{Steps (Round 0)}:
\begin{enumerate}[leftmargin=*,noitemsep]
    \item Explore the repo and read the source files mentioned in the issue. Docstrings and comments often reveal the intended contract.
    \item \textbf{Boundary localization (REQUIRED)}: produce (a)~a boundary chain from public API to symptom, (b)~the chosen boundary and observable your reproduction will exercise, (c)~two plausible candidates for where the gold fix might live, and (d)~an assertion derivation that justifies each assertion by observation, project specification, or buggy-code output. Walk through $P$ on buggy, fixed, and unrelated-crash states before committing.
    \item Create \texttt{reproduction.py} inside the project directory using the chosen boundary. Trigger the bug the same way an end-user would: invoke the public CLI, call the documented public API, or hit the HTTP endpoint the reporter used. Do not substitute the project's test-infrastructure helpers.
    \item Execute \texttt{python reproduction.py} and confirm it exits non-zero because of the bug.
    \item Call \texttt{finish} --- you will receive feedback guiding you through the remaining steps.
\end{enumerate}

\smallskip
\textbf{Conventions}: Create \texttt{reproduction.py} inside \texttt{/workspace/\{\{ workspace\_dir\_name \}\}/}. Keep all changes uncommitted throughout the process --- the detector evaluates via \texttt{git diff}. Use \texttt{git stash} / \texttt{git stash pop} for F$\to$P verification.

Your thinking should be thorough and so it's fine if it's very long.
\end{tcolorbox}
\newpage


\begin{tcolorbox}[
  breakable,
  enhanced,
  colback=lightgray!20,
  colframe=darkgray!80,
  title=\ours{} Bootstrap Prompt (Round 1)
]
\label{app:prompt_bootstrap}
\small

Round 0 complete --- your \texttt{reproduction.py} correctly detects the bug (exits non-zero on buggy code).

\smallskip
\textbf{What to do next (Round 1: identify scope, find the boundary, write the fix)}:

\begin{enumerate}[leftmargin=*,noitemsep]
    \item \textbf{Identify scope (sibling enumeration)} --- bugs often manifest at multiple sibling-shaped sites that \texttt{reproduction.py} exercises only one of. Before debugging, enumerate the candidate scope. Sibling shapes to consider:
    \begin{itemize}[leftmargin=*,noitemsep]
        \item Parallel implementations: per-backend, per-renderer, subclass overrides
        \item Mirror printers/converters: \texttt{str/latex/pretty/repr}, \texttt{to\_python/from\_python}, \texttt{encode/decode}
        \item Caller direction: who CALLS the buggy function?
        \item Companion methods: paired changes (\texttt{add} $\leftrightarrow$ \texttt{clear}, \texttt{\_\_mul\_\_} $\leftrightarrow$ \texttt{\_\_rmul\_\_})
    \end{itemize}
    For each candidate shape, run \texttt{grep -rln '<symbol>' --include='*.py'} and inspect each result. Record the sibling files --- they are the SCOPE your fix must address.

    \item \textbf{Debug to find the root cause}: Starting from the symptom your \texttt{reproduction.py} exhibits, trace inward along the witness chain to the \textbf{boundary} --- the first inner layer where the behavior turns from correct to wrong.

    \item \textbf{Determine expected behavior}: Based on your root cause analysis, form a hypothesis about what the correct behavior should be. Verify by examining the project's specification (existing tests, docstrings, API contracts).

    \item \textbf{Strengthen \texttt{reproduction.py} along the boundary}: state the expected vs actual behavior at the boundary, then add hypothesis checks --- assertions on intermediate values along the boundary path. If step~1 surfaced multiple sibling sites, add at least one assertion per sibling.

    \item \textbf{Write a fix} targeting the boundary --- each change should follow from the expected behavior in step~3. Apply the fix at every in-scope sibling that shares the buggy pattern. Run the entire test module to check for regressions:
    \texttt{PYTHONPATH=\{\{ repo\_dir \}\} \{\{ test\_framework \}\} <test\_module>}

    \item \textbf{Verify F$\to$P}: Revert only your fix (not the test) by running \texttt{git stash push <fix\_files>}, run \texttt{reproduction.py} to confirm it FAILS, then restore with \texttt{git stash pop}.

    \item Call \texttt{finish}.
\end{enumerate}

\end{tcolorbox}


\begin{tcolorbox}[
  breakable,
  enhanced,
  colback=lightgray!20,
  colframe=darkgray!80,
  title=\ours{} Hardening Feedback Prompt (Round $\geq 2$)
]
\label{app:prompt_hardening}
\small

\texttt{\{\{ n\_total $-$ n\_rejected \}\}/\{\{ n\_total \}\}} mutants accepted. Vs prev round: \texttt{\{\{ temporal.a\_T \}\}} persistent $\cdot$ \texttt{\{\{ temporal.b\_T \}\}} regressed $\cdot$ \texttt{\{\{ temporal.c\_T \}\}} newly killed $\checkmark$.

\smallskip
\textbf{Why this matters}: This mutant slipped through your test, which means your test is narrow on some dimension --- it cannot tell your fix apart from a wrong-fix that produces the same symptom. By symmetry, your fix may have the same blind spot. Re-debug from the symptom; then strengthen BOTH \texttt{reproduction.py} and the fix.

\smallskip
\textbf{Top accepted mutant}:

\smallskip
\texttt{\{\{ top\_exemplar\_diff \}\}}

\smallskip
Other surviving operators: \texttt{\{\{ surviving\_ops[1:] | join(', ') \}\}}. Per-mutant diffs + hints in \texttt{/workspace/.adversarial\_feedback/latest.md}.

\smallskip
\textbf{Fix-side audit (run before \texttt{finish})}:
\begin{enumerate}[leftmargin=*,noitemsep]
    \item \textbf{Root-cause depth}: is the function you edited the deepest one returning the wrong value? If the symptom flows from a callee whose value is also wrong, fix the callee --- not the symptom site.
    \item \textbf{Sibling-fix enumeration}: for each symbol your fix touches, run \texttt{grep -rln '<symbol>'} and inspect each non-edited file. Apply the fix at every site sharing the buggy pattern, or articulate why each is unaffected.
    \item \textbf{Scope check}: every file in your diff should appear in the issue's stack trace or be named in the issue text.
    \item \textbf{Output-shape match}: if the fix changes a returned value, message, or exception, predict the exact output and compare to the issue text's expected output. If your prediction differs, restructure the fix.
\end{enumerate}

\smallskip
\textbf{Verify before \texttt{finish}}: (1)~Re-run \texttt{reproduction.py} $\to$ F$\to$P holds. (2)~Run existing tests for your edited symbol; any newly-failing test $\to$ narrow the fix.

\end{tcolorbox}

\newpage


\begin{tcolorbox}[
  breakable,
  enhanced,
  colback=lightgray!20,
  colframe=darkgray!80,
  title=\fixonly{} Prompt
]
\label{app:prompt_fixonly}
\small

I have access to a python code repository in the directory \texttt{\{\{ instance.repo\_path \}\}}. You can explore and modify files using the available tools. Consider the following issue description:

\smallskip
\texttt{<issue\_description>}\\
\texttt{\{\{ instance.problem\_statement \}\}}\\
\texttt{</issue\_description>}

\smallskip
Can you help me implement the necessary changes to the repository so that the requirements specified in the \texttt{<issue\_description>} are met?
I've already taken care of all changes to any of the test files described in the \texttt{<issue\_description>}. This means you DON'T have to modify the testing logic or any of the tests in any way!
Also the development Python environment is already set up for you (i.e., all dependencies already installed), so you don't need to install other packages.
Your task is to make the minimal changes to non-test files in the \texttt{\{\{ instance.repo\_path \}\}} directory to ensure the \texttt{<issue\_description>} is satisfied.

\smallskip
Follow these phases to resolve the issue:

\smallskip
\textbf{Phase 1. READING}: read the problem and reword it in clearer terms
\begin{itemize}[leftmargin=*,noitemsep]
    \item 1.1 If there are code or config snippets, express in words any best practices or conventions in them.
    \item 1.2 Highlight message errors, method names, variables, file names, stack traces, and technical details.
    \item 1.3 Explain the problem in clear terms.
    \item 1.4 Enumerate the steps to reproduce the problem.
    \item 1.5 Highlight any best practices to take into account when testing and fixing the issue.
\end{itemize}

\smallskip
\textbf{Phase 2. RUNNING}: install and run the tests on the repository
\begin{itemize}[leftmargin=*,noitemsep]
    \item 2.1 Activate the environment by running \texttt{./opt/miniconda3/etc/profile.d/conda.sh ; conda activate testbed}
    \item 2.2 Follow the readme and install the environment and anything needed.
    \item 2.3 Iterate and figure out how to run the tests.
\end{itemize}

\smallskip
\textbf{Phase 3. EXPLORATION}: find the files that are related to the problem and possible solutions
\begin{itemize}[leftmargin=*,noitemsep]
    \item 3.1 Use \texttt{grep} to search for relevant methods, classes, keywords and error messages.
    \item 3.2 Identify all files related to the problem statement.
    \item 3.3 Propose the methods and files to fix the issue and explain why.
    \item 3.4 From the possible file locations, select the most likely location to fix the issue.
\end{itemize}

\smallskip
\textbf{Phase 4. TEST CREATION}:  create a script to reproduce and verify the issue
\begin{itemize}[leftmargin=*,noitemsep]
    \item 4.1 Look at existing test files in the repository to understand the test format/structure.
    \item 4.2 Create a minimal reproduction script that reproduces the located issue.
    \item 4.3 Run the reproduction script to confirm you are reproducing the issue.
    \item 4.4 Adjust the reproduction script as necessary.
\end{itemize}

\smallskip
\textbf{Phase 5. FIX ANALYSIS}: state clearly the problem and how to fix it
\begin{itemize}[leftmargin=*,noitemsep]
    \item 5.1 State clearly what the problem is.
    \item 5.2 State clearly where the problem is located.
    \item 5.3 State clearly how the test reproduces the issue.
    \item 5.4 State clearly the best practices to take into account in the fix.
    \item 5.5 State clearly how to fix the problem.
\end{itemize}

\smallskip
\textbf{Phase 6. FIX IMPLEMENTATION}: Edit the source code to implement your chosen solution.
\begin{itemize}[leftmargin=*,noitemsep]
    \item 6.1 Make minimal, focused changes to fix the issue.
\end{itemize}

\smallskip
\textbf{Phase 7. VERIFICATION}: Test your implementation thoroughly.
\begin{itemize}[leftmargin=*,noitemsep]
    \item 7.1 Run your reproduction script to verify the fix works.
    \item 7.2 Add edge cases to your test script to ensure comprehensive coverage.
    \item 7.3 Run existing tests related to the modified code to ensure you haven't broken anything.
\end{itemize}

\smallskip
\textbf{Phase 8. FINAL REVIEW}: Carefully re-read the problem description and compare your changes with the base commit \texttt{\{\{ instance.base\_commit \}\}}.
\begin{itemize}[leftmargin=*,noitemsep]
    \item 8.1 Ensure you've fully addressed all requirements.
    \item 8.2 Run any tests in the repository related to: the issue you are fixing, the files you modified, and the functions you changed.
    \item 8.3 If any tests fail, revise your implementation until all tests pass.
\end{itemize}
Be thorough in your exploration, testing, and reasoning. It's fine if your thinking process is lengthy - quality and completeness are more important than brevity.
\end{tcolorbox}


\begin{tcolorbox}[
  breakable,
  enhanced,
  colback=lightgray!20,
  colframe=darkgray!80,
  title=TDD-style BRT Injection Prompt
]
\label{app:prompt_tdd}
\small

I have access to a python code repository in the directory \texttt{\{\{ instance.repo\_path \}\}}. You can explore and modify files using the available tools. Consider the following issue description:

\smallskip
\texttt{<issue\_description>}\\
\texttt{\{\{ instance.problem\_statement \}\}}\\
\texttt{</issue\_description>}

\smallskip
A reproduction test has already been written for this issue. Apply the test below to the repository, then fix the source code so the test passes.

\smallskip
\texttt{<reproduction\_test\_patch>}\\
\texttt{\{\{ instance.tdd\_test\_patch \}\}}\\
\texttt{</reproduction\_test\_patch>}

\smallskip
The development Python environment is already set up for you (i.e., all dependencies already installed), so you don't need to install other packages.
Your task is to make the minimal changes to non-test files in the \texttt{\{\{ instance.repo\_path \}\}} directory to ensure the \texttt{<issue\_description>} is satisfied and the reproduction test passes.

\smallskip
Follow these phases to resolve the issue:

\smallskip
\textbf{Phase 1. READING}: read the problem and the reproduction test
\begin{itemize}[leftmargin=*,noitemsep]
    \item 1.1 Read the issue description and reword it in clearer terms.
    \item 1.2 Read the reproduction test patch carefully. Identify which test file(s) and test function(s) it adds or modifies.
    \item 1.3 Understand what behavior the test asserts --- what inputs does it use, what outputs does it expect?
    \item 1.4 Explain the connection between the issue and the test: how does the test expose the bug?
\end{itemize}

\smallskip
\textbf{Phase 2. TEST SETUP}: apply the reproduction test to the repository
\begin{itemize}[leftmargin=*,noitemsep]
    \item 2.1 Activate the environment by running \texttt{./opt/miniconda3/etc/profile.d/conda.sh ; conda activate testbed}
    \item 2.2 Apply the reproduction test patch using \texttt{git apply} (or manually create the test files).
    \item 2.3 Run the reproduction test to confirm it \textbf{fails} on the current (buggy) code.
    \item 2.4 If the test does not fail, re-read the patch and the issue --- you may need to adjust how you apply it.
\end{itemize}

\smallskip
\textbf{Phase 3. EXPLORATION}: find the source code related to the failing test
\begin{itemize}[leftmargin=*,noitemsep]
    \item 3.1 Use \texttt{grep} to search for the methods, classes, and keywords referenced by the test and the issue.
    \item 3.2 Identify all source files related to the problem.
    \item 3.3 Propose candidate fix locations and explain why.
    \item 3.4 Select the most likely location to fix.
\end{itemize}

\smallskip
\textbf{Phase 4. FIX ANALYSIS}: plan the fix
\begin{itemize}[leftmargin=*,noitemsep]
    \item 4.1 State clearly what the problem is.
    \item 4.2 State clearly where the problem is located.
    \item 4.3 State clearly how the reproduction test exposes the issue.
    \item 4.4 State clearly the best practices to take into account in the fix.
    \item 4.5 State clearly how to fix the problem.
\end{itemize}

\smallskip
\textbf{Phase 5. FIX IMPLEMENTATION}: Edit the source code to implement your chosen solution.
\begin{itemize}[leftmargin=*,noitemsep]
    \item 5.1 Make minimal, focused changes to fix the issue.
\end{itemize}

\smallskip
\textbf{Phase 6. VERIFICATION}: Test your implementation thoroughly.
\begin{itemize}[leftmargin=*,noitemsep]
    \item 6.1 Run the reproduction test to verify the fix makes it pass.
    \item 6.2 Run existing tests related to the modified code to ensure you haven't broken anything.
    \item 6.3 If any tests fail, revise your implementation until all tests pass.
\end{itemize}

\smallskip
\textbf{Phase 7. FINAL REVIEW}: Carefully re-read the problem description and compare your changes with the base commit \texttt{\{\{ instance.base\_commit \}\}}.
\begin{itemize}[leftmargin=*,noitemsep]
    \item 7.1 Ensure you've fully addressed all requirements.
    \item 7.2 Run any tests in the repository related to: the issue you are fixing, the files you modified, and the functions you changed.
    \item 7.3 If any tests fail, revise your implementation until all tests pass.
\end{itemize}
Be thorough in your exploration, testing, and reasoning. It's fine if your thinking process is lengthy - quality and completeness are more important than brevity.
\end{tcolorbox}

\newpage


\begin{tcolorbox}[
  breakable,
  enhanced,
  colback=lightgray!20,
  colframe=darkgray!80,
  title=\cogen{} Prompt
]
\label{app:prompt_cogen}
\small

\texttt{<uploaded\_files>}\\
\texttt{/workspace/\{\{ workspace\_dir\_name \}\}}\\
\texttt{</uploaded\_files>}

\smallskip
I've uploaded a python code repository in the directory \texttt{\{\{ workspace\_dir\_name \}\}}. Consider the following issue description:

\smallskip
\texttt{<issue\_description>}\\
\texttt{\{\{ instance.problem\_statement \}\}}\\
\texttt{</issue\_description>}

\smallskip
\{\% block task \%\}\\
Your task is to fix this bug and write a test that reproduces it. Both the fix and the test must be in your final patch.\\
\{\% endblock \%\}

\smallskip
Your fix should resolve the issue with minimal changes. Your test should reproduce the issue --- i.e., it should fail before the fix is applied and pass after. Make sure the test is implementation-agnostic: it should verify the correct behavior described in the issue, not details specific to your fix.

\smallskip
\textbf{Important}: After you submit, your test will be automatically evaluated against alternative code patches --- some may be correct fixes using a different approach, others may be plausible but subtly wrong. Your test should be robust enough to:
\begin{itemize}[leftmargin=*,noitemsep]
    \item PASS on any correct fix (regardless of implementation details)
    \item FAIL on patches that don't fully resolve the underlying bug
\end{itemize}
Use your understanding of the issue semantics to decide what the correct behavior should be, and test for that rather than for artifacts of your particular fix.

\smallskip
Follow these steps to resolve the issue:
\begin{enumerate}[leftmargin=*,noitemsep]
    \item As a first step, it might be a good idea to explore the repo to familiarize yourself with its structure.
    \item Create a script \texttt{reproduction.py} to reproduce the error and execute it with \texttt{python reproduction.py} using the BashTool, to confirm the error.
    \item The issue describes a SYMPTOM. Find WHERE the root cause is --- it may be in a different module than where the symptom appears. Test one hypothesis at a time.
    \item Fix the bug with a minimal code change.
    \item Re-run \texttt{reproduction.py} to confirm the fix resolves the issue.
    \item Run the existing test suite for the module(s) you modified to make sure your fix doesn't break anything. Fix any regressions before proceeding.
    \item Edit the sourcecode of the repo to integrate your reproduction script into the test framework. Find the most relevant existing test file. Make sure your test fails on the buggy code and passes on the fixed code.
    \item Verify F$\to$P: revert your fix with \texttt{git stash}, run your test to confirm it FAILS, then restore with \texttt{git stash pop}.
    \item Clean up: remove \texttt{reproduction.py} and any other scratch files before finishing. Only source and test file changes should remain in your patch.
\end{enumerate}

\end{tcolorbox}

\newpage

\begin{tcolorbox}[
  breakable,
  enhanced,
  colback=lightgray!20,
  colframe=darkgray!80,
  title=\brtonly{} Prompt
]
\label{app:prompt_brtonly}
\small

\texttt{<uploaded\_files>}\\
\texttt{/workspace/\{\{ workspace\_dir\_name \}\}}\\
\texttt{</uploaded\_files>}

\smallskip
I've uploaded a python code repository in the directory \texttt{\{\{ workspace\_dir\_name \}\}}. Consider the following issue description:

\smallskip
\texttt{<issue\_description>}\\
\texttt{\{\{ instance.problem\_statement \}\}}\\
\texttt{</issue\_description>}

\smallskip
Can you help me implement the necessary changes to the repository to test whether the issue in \texttt{<issue\_description>} was resolved?
I will take care of all changes to any of the non-test files. This means you DON'T have to modify the actual logic and ONLY have to update test logic and tests!

\smallskip
Your task is to make the minimal changes to tests files in the \texttt{/workspace} directory to reproduce the issue in the \texttt{<issue\_description>}, i.e., such that the generated tests fail in the current state (where the issue is unresolved) and pass when the issue will be resolved.

\smallskip
Follow these steps to reproduce the issue:
\begin{enumerate}[leftmargin=*,noitemsep]
    \item As a first step, it might be a good idea to explore the repo to familiarize yourself with its structure.
    \item Create a script \texttt{reproduction.py} to reproduce the error and execute it with \texttt{python reproduction.py} using the BashTool, to confirm the error.
    \item Edit the sourcecode of the repo to integrate your reproduction script into the test framework.
    \item Run the test framework and make sure your tests fail! Only submit FAILING tests! Never submit passing tests.
\end{enumerate}
Your thinking should be thorough and so it's fine if it's very long.
\end{tcolorbox}


%

\end{document}